\documentclass[11pt]{article}
\usepackage{latexsym,amssymb,amstext}
\def\slash#1{\rlap{\hbox{$\mskip 1 mu /$}}#1}      
\def\Slash#1{\rlap{\hbox{$\mskip 3 mu /$}}#1}      
\newcommand{\ft}[2]{{\textstyle\frac{#1}{#2}}}

\begin{document} 
%
\begin{titlepage}
\begin{flushright} \small
 ITP-UU-07/35 \\ SPIN-07/24 
\end{flushright}
\bigskip

\begin{center}
 {\LARGE\bfseries Lagrangians with electric and magnetic charges
   \\[3mm] 
   of N=2  supersymmetric gauge theories}  
\\[10mm]
\textbf{Mathijs de Vroome and Bernard de Wit}\\[5mm]
{\em Institute for Theoretical Physics} and {\em Spinoza
  Institute,\\ Utrecht University, Utrecht, The Netherlands}\\[3mm] 
{\tt M.T.deVroome@phys.uu.nl}\quad,\quad {\tt  B.deWit@phys.uu.nl}
\end{center}

\vspace{5mm}

\bigskip

\centerline{\bfseries Abstract} 
\medskip
\noindent
General Lagrangians are constructed for N=2 supersymmetric gauge
theories in four space-time dimensions involving gauge groups with
(non-abelian) electric and magnetic charges. The charges induce a
scalar potential, which, when the charges are regarded as spurionic
quantities, is invariant under electric/magnetic duality. The
resulting theories are especially relevant for supergravity, but
details of the extension to local supersymmetry will be discussed
elsewhere. The results include the coupling to hypermultiplets.
Without the latter, it is demonstrated how an off-shell representation
can be constructed based on vector and tensor supermultiplets.
\medskip
\end{titlepage}

\section{Introduction}
\label{sec:introduction}
In four space-time dimensions, theories with abelian gauge fields may
have more symmetries than are apparent from the Lagrangian (or the
corresponding action). The full invariance group may include
symmetries of the combined field equations and Bianchi identities that
are not realized at the level of the Lagrangian. This group is a
subgroup of the electric/magnetic duality group, which, for $n$ vector
fields, is equal to $\mathrm{Sp}(2n,\mathbb{R})$.  Under a generic
electric/magnetic duality the Lagrangian will in general change, but
the new Lagrangian will still lead to an equivalent set of field
equations and Bianchi identities.  Therefore these different
Lagrangians, which do not have to share the same symmetry group,
belong to the same equivalence class.  When the Lagrangian does not
change under a duality (possibly after combining with corresponding
transformations of the other fields) one is dealing with an invariance
of the theory. To appreciate this feature, it is important to note
that a Lagrangian does not transform as a function under duality
transformations. In fact the gauge fields before and after the
transformation are not related by a local field redefinition. This is
the underlying reason why the full invariance is not necessarily
reflected by an invariance of the Lagrangian that is induced by
transformations of the various fields.

When introducing charges for some of the fields, the standard
procedure is to introduce minimal couplings and covariant field
strengths in the Lagrangian. This implies that the charges are all
electric. The gauge group will therefore be contained in the
invariance group of the Lagrangian, so that one cannot necessarily
gauge any subgroup of the full invariance group. In that case one has
two options. Either one uses electric/magnetic duality to obtain
another Lagrangian belonging to the same equivalence class that has a
more suitable invariance group in which the desired gauge group can be
embedded, or, one uses a recently proposed formalism that incorporates
both electric and magnetic charges \cite{deWit:2005ub}. The latter
allows one to start from any particular Lagrangian belonging to a
certain equivalence class, provided that this class contains at least
one Lagrangian in which all the charges that one intends to switch on
are electric.

In this paper we study general gaugings of $N=2$ supersymmetric gauge
theories, based on vector multiplets and hypermultiplets. It is well
known that the introduction of charged fields in a supersymmetric
field theory tends to break supersymmetry.  To preserve supersymmetry
the theory has to be extended with a scalar potential and masslike
terms. The goal is to derive these terms in the context of the
formalism presented in \cite{deWit:2005ub}. It is not the first time
that this formalism has been used for four-dimensional supersymmetric
theories. In \cite{Schon:2006kz} it was successfully applied to $N=4$
supergravity and in \cite{deWit:2007mt} to $N=8$ supergravity. In this
approach the cumbersome procedure according to which the ungauged
Lagrangian has to be converted to a suitable electric frame, prior to
switching on the charges, is avoided.  Moreover, the scalar potential
and masslike terms that accompany the gaugings are found in a way that
is independent of the electric/magnetic duality frame. By introducing
both electric {\it and} magnetic charges the potential will thus fully
exhibit the duality invariances. This is of interest, for example,
when studying flux compactifications in string theory, because the
underlying fluxes are usually subject to integer-valued rotations
associated to the non-trivial cycles of the underlying internal
manifold.

The framework of \cite{deWit:2005ub} incorporates both electric and
magnetic charges and their corresponding gauge fields.  The charges
are encoded in terms of a so-called embedding tensor, which defines
the embedding of the gauge group into the full rigid invariance group.
This embedding tensor is treated as a spurionic object, so that the
electric/magnetic duality structure of the ungauged theory is
preserved after charges are turned on. Besides introducing a set of
dual magnetic gauge fields, tensor fields are required that transform
in the adjoint representation of the rigid invariance group. These
extra fields carry additional off-shell degrees of freedom, but the
number of physical degrees of freedom remains the same, owing to extra
gauge transformations.  Prior to \cite{deWit:2005ub} it had already
been discovered that magnetic charges tend to be accompanied by tensor
fields. An early example of this was presented in \cite{Louis:2002ny},
and subsequently more theories with magnetic charges and tensor fields
were constructed, for instance, in
\cite{Dall'Agata:2003yr,Sommovigo:2004vj,DAuria:2004yi}.  However, in
these references the gauge groups are abelian.

The starting point of this paper is the expression for $N=2$
supersymmetric Lagrangians of $n$ vector supermultiplets, labeled by
indices $\Lambda=1,\ldots,n$. This Lagrangian is encoded in terms of a
holomorphic function $F(X)$, which, for the abelian case, takes the
following form,
\begin{eqnarray}
  \label{eq:L-0}
  \mathcal{L}_0 &=& 
  \mathrm{i}\, \partial_\mu F_\Lambda\,\partial^\mu \bar X^\Lambda
  + \ft12 \mathrm{i}\,F_{\Lambda\Sigma} \, \bar\Omega_i{}^\Lambda
  \slash{\partial} \Omega^{i\Sigma} 
  + \ft14 \mathrm{i}\, F_{\Lambda\Sigma}\, F^-_{\mu\nu}{}^\Lambda\,
  F^{-\Sigma\,\mu\nu} 
  - \ft{1}{8}\mathrm{i} \,F_{\Lambda\Sigma}\, Y_{ij}{}^\Lambda \,
  Y^{ij\Sigma}  
  \nonumber\\
  &&
  + \ft{1}{8} \mathrm{i}\,  F_{\Lambda\Sigma\Gamma}\,
  Y^{ij\Lambda}\, \bar{\Omega}_{i}^{\Sigma} \Omega_{j}^{\Gamma} 
  -\ft1{16} \mathrm{i} F_{\Lambda\Sigma\Gamma}\,
  \bar\Omega_i{}^\Lambda\gamma^{\mu\nu} \Omega _j^\Gamma\,
  \varepsilon^{ij} \,F_{\mu\nu}^-{}^{\Sigma}\, 
  \nonumber\\
   && 
   - \ft{1}{48} \mathrm{i} \,\varepsilon^{ij} \varepsilon^{kl} \,
    F_{\Lambda\Sigma\Gamma\Xi}\, \bar{\Omega}_i^{\Lambda}
   \Omega_k^{\Sigma} \; \bar{\Omega}_j^{\Gamma} \Omega_l^{\Xi} 
    \nonumber \\
      &&
      + \mathrm{h.c.} \;, 
\end{eqnarray}
where $F_{\Lambda_1\cdots \Lambda_k}$ denotes the $k$-th derivative of
$F(X)$. The fermion fields $\Omega^\Lambda$ and the auxiliary fields
$Y^\Lambda$ carry $\mathrm{SU}(2)$ indices $i,j, \ldots=1,2$. Spinors
$\Omega_i{}^\Lambda$ have positive, and spinors $\Omega^{i\Lambda}$ have
negative chirality (so that $\gamma^5 \Omega_i{}^\Lambda=
\Omega_i{}^\Lambda$ and $\gamma^5 \Omega^{i\Lambda}=
-\Omega^{i\Lambda}$). The auxiliary fields satisfy the pseudo-reality
constraint $(Y_{ij}{}^\Lambda)^\ast = \varepsilon^{ik}\varepsilon^{jl}
Y_{kl}{}^\Lambda$. The tensors $F^{\pm}_{\mu\nu}{}^\Lambda$ are the
(anti-)selfdual components of the field strengths, which will be
expressed in terms of vector fields $A_\mu{}^\Lambda$.  Even when all
charges are electric it is possible that the function $F(X)$ is not
invariant under the gauge group. In that case the gauge group must be
non-semisimple \cite{deWit:1984px}. The gauge group for the
hypermultiplets can be either abelian or non-abelian, but a
non-trivial gauge group for the vector multiplets is always
non-abelian, possibly with a central extension.

The supersymmetric Lagrangians derived in this paper incorporate
gaugings in both the vector and hypermultiplet sectors. Although the
vector multiplets are originally defined as off-shell multiplets, the
presence of the magnetic charges causes a breakdown of off-shell
supersymmetry. Of course, hypermultiplets are not based on an
off-shell representation of the supersymmetry algebra irrespective of
the presence of charges. It is an interesting question whether the
results of this chapter can be reformulated such that the vector
multiplets retain their off-shell form and, indeed, we show that such
an off-shell version can be constructed based on vector and tensor
supermultiplets. However, we refrain from considering the
extension of the theories of this paper to supergravity. This extension
is expected to be straightforward upon use of the superconformal
multiplet calculus \cite{deWit:1980tn,deWit:1984pk,deWit:1984px}. We
intend to return to this topic elsewhere.

This paper is organized as follows. In section
\ref{vector-multiplets-duality} we recall the relevant features of
$N=2$ vector multiplets and electric/magnetic duality, and discuss the
introduction of electric and magnetic charges. In section
\ref{sec:embedding-tensor} we introduce the embedding tensor and we
review the formalism of \cite{deWit:2005ub}.  Section
\ref{sec:rest-supersymm-non} deals with the restoration of
supersymmetry in vector multiplet models after a gauging, and section
\ref{sec:hypermultiplets} gives the extension with hypermultiplets.
The off-shell formulation of the theories of this paper is discussed
in section \ref{sec:off-shell-structure}, and in section
\ref{sec:summary-discussion} we summarize the results obtained and
briefly indicate some of their applications.

\section{Vector multiplets, electric/magnetic duality, and non-abelian
  charges} 
\label{vector-multiplets-duality} 
\setcounter{equation}{0} 
In this section we discuss electric/magnetic duality and the
introduction of charges for systems of vector supermultiplets. To
facilitate the presentation it is convenient to decompose the
Lagrangian (\ref{eq:L-0}) as follows,
\begin{equation}
  \label{eq:L-vector}
  \mathcal{L}_0 = \mathcal{L}_{\mathrm{matter}} + \mathcal{L}_{\mathrm{kin}} 
  + \mathcal{L}_{\Omega^4} + \mathcal{L}_Y\,,
\end{equation}
where $\mathcal{L}_{\mathrm{matter}}$ contains the kinetic terms
of the scalar and spinor fields,
\begin{eqnarray}
  \label{eq:L-matter}
  \mathcal{L}_{\mathrm{matter}}  &=&
  \mathrm{i}\Big(\partial_\mu F_\Lambda\,\partial^\mu \bar X^\Lambda
  -\partial_\mu \bar F_\Lambda\,\partial^\mu  X^\Lambda \Big) 
  \nonumber\\ 
  & & 
  -\ft14 N_{\Lambda\Sigma}\Big(\bar\Omega^{i\Lambda}\slash{\partial} 
  \Omega _i{}^\Sigma 
      +\bar \Omega_i{}^\Lambda \slash{\partial}\Omega ^{i\Sigma}\Big)
    -\ft14 \mathrm{i}\Big(\bar \Omega_i{}^\Lambda\slash{\partial} 
    F_{\Lambda\Sigma} \Omega ^{i\Sigma} 
    -\bar\Omega^{i \Lambda} \slash{\partial} \bar F_{\Lambda\Sigma}
  \Omega _i{}^\Sigma\Big) \;. 
\end{eqnarray}
The kinetic terms of the vector fields combined with a number of terms
that are related to them by electric/magnetic duality, are contained in
$\mathcal{L}_{\mathrm{vector}}$,
\begin{eqnarray}
  \label{eq:vlagr-kin}
  \mathcal{L}_{\mathrm{vector}}  &=&
  \ft14 \mathrm{i}\Big(F_{\Lambda\Sigma}F^{-\Lambda}_{\mu\nu}
  F^{-\Sigma\,\mu\nu}
  -\bar F_{\Lambda\Sigma}F^{+\Lambda}_{\mu\nu}F^{+\Sigma\,\mu\nu}\Big)
  \nonumber\\
  && 
  -\ft1{16} \mathrm{i} \Big( F_{\Lambda\Sigma\Gamma}\bar\Omega
  _i^\Lambda\,\gamma^{\mu\nu}  
  F_{\mu\nu}^{-\Sigma}\,\Omega _j^\Gamma\, \varepsilon^{ij}
      -\bar F_{\Lambda\Sigma\Gamma}\bar \Omega ^{i\Lambda}\,\gamma^{\mu\nu} 
      F_{\mu\nu}^{+\Sigma}\,\Omega ^{j\Gamma}\,\varepsilon_{ij} \Big) 
      \nonumber\\
  &&
  - \ft1{256}\mathrm{i} N^{\Delta\Omega} \Big(F_{\Delta\Lambda\Sigma}
      \bar\Omega _i{}^\Lambda\gamma_{\mu\nu}\Omega _j{}^\Sigma
  \varepsilon^{ij} \Big) \Big(F_{\Gamma\Xi\Omega} 
  \bar\Omega _k{}^\Gamma\gamma^{\mu\nu}\Omega _l{}^\Xi
  \varepsilon^{kl}\Big) 
    \nonumber\\
  &&
  + \ft1{256}\mathrm{i} N^{\Delta\Omega} \Big(\bar F_{\Delta\Lambda\Sigma}
  \bar\Omega ^{i\Lambda}\gamma_{\mu\nu}\Omega ^{j\Sigma}
  \varepsilon_{ij} \Big)  \Big(\bar F_{\Gamma\Xi\Omega} 
  \bar\Omega ^{k\Gamma}\gamma^{\mu\nu}\Omega ^{l\Xi}
  \varepsilon_{kl}\Big)  \;. 
\end{eqnarray}
Quartic spinor terms that are consistent with respect to
electric/magnetic duality, are given by 
\begin{eqnarray}
  \label{eq:vlagr-4}
  \mathcal{L}_{\Omega^4} &=& 
  \ft1{384}\mathrm{i} \Big(F_{\Lambda\Sigma\Gamma\Xi} 
  +3\, \mathrm{i} N^{\Delta\Omega} \,F_{\Delta(\Lambda\Gamma}
  F_{\Sigma\Xi)\Omega} \Big) \,  
  \bar\Omega _i{}^\Lambda\gamma_{\mu\nu}\Omega _j{}^\Sigma \varepsilon^{ij} \,
  \bar\Omega _k{}^\Gamma\gamma^{\mu\nu}\Omega _l{}^\Xi \varepsilon^{kl}
  \nonumber \\ 
  && 
  -\ft1{384}\mathrm{i} \Big(\bar F_{\Lambda\Sigma\Gamma\Xi} 
  -3\, \mathrm{i} N^{\Delta\Omega}\bar F_{\Delta(\Lambda\Gamma}\bar
  F_{\Sigma \Xi)\Omega} \Big) \,  
  \bar\Omega ^{i\Lambda}\gamma_{\mu\nu}\Omega^{j\Sigma} \varepsilon_{ij} \,
  \bar\Omega ^{k\Gamma}\gamma^{\mu\nu}\Omega^{l\Xi} \varepsilon_{kl}
  \nonumber\\  
  && 
  -\ft1{16} N^{\Delta\Omega} F_{\Delta\Lambda\Sigma} \bar
  F_{\Gamma\Xi\Omega}\,\bar\Omega^{i\Gamma}\Omega ^{j\Xi}\, 
     \bar\Omega _i{}^\Lambda\Omega _j{}^\Sigma\,,
\end{eqnarray}
and, finally, $\mathcal{L}_Y$ comprises the terms associated with the
auxiliary fields $Y_{ij}{}^\Lambda$,
\begin{eqnarray}
  \label{eq:vlagr-Y}
     \mathcal{L}_Y&=& \ft1{8}
     N^{\Lambda\Sigma}\,\left (N_{\Lambda\Gamma}Y_{ij}{}^\Gamma +
     \ft12\mathrm{i} (F_{\Lambda\Gamma\Omega}\,
     \bar\Omega_i{}^\Gamma\Omega_j{}^\Omega - \bar F_{\Lambda\Gamma\Omega} 
     \,\bar\Omega^{k\Gamma}\Omega^{l\Omega}\varepsilon_{ik}\varepsilon_{jl}
     )\right)  \nonumber\\ 
     && {}
    \times\left(N_{\Sigma\Xi}Y^{ij\Xi}  +
     \ft12\mathrm{i} (F_{\Sigma\Xi\Delta}
     \,\bar\Omega_m{}^\Xi\Omega_n{}^\Delta
     \varepsilon^{im}\varepsilon^{jn} -  \bar F_{\Sigma\Xi\Delta} 
     \,\bar\Omega^{i\Xi}\Omega^{j\Delta} )  \right)\,. 
\end{eqnarray} 
This last result for $\mathcal{L}_Y$ is not obviously consistent with
electric/magnetic duality. We return to this in a sequal. Here and
henceforth we use the notation,
\begin{equation}
  \label{eq:def-N}
  N_{\Lambda\Sigma} = -\mathrm{i} F_{\Lambda\Sigma} + \mathrm{i}\bar
  F_{\Lambda\Sigma} \,,  
  \qquad 
  N^{\Lambda\Sigma}\equiv \big[N^{-1}\big]^{\Lambda\Sigma}\,.
\end{equation}
Note that $N_{\Lambda\Sigma}$ plays the role of the inverse effective
coupling constants while the real part of $F_{\Lambda\Sigma}$ plays
the role of the generalized theta angles. 

The non-linear sigma model contained in (\ref{eq:vlagr-kin}) exhibits
an interesting geometry known as {\it special geometry}. The complex
scalars $X^\Lambda$ parametrize an $n$-dimensional target space with
metric $g_{\Lambda\bar\Sigma}= N_{\Lambda\Sigma}$. This is a K\"ahler
space: its metric equals $g_{\Lambda\bar\Sigma} = \partial^2 K(X,\bar
X)/\partial X^\Lambda\,\partial \bar X^\Sigma$, with K\"ahler
potential
\begin{equation}
  \label{eq:Kpotential}
  K(X,\bar X) = \mathrm{i} X^\Lambda\, \bar F_\Lambda(\bar X) -
  \mathrm{i}\bar X^\Lambda \,  F_\Lambda(X) \,.  
\end{equation}

The supersymmetry transformations that leave the action corresponding
to (\ref{eq:L-vector}) invariant, are given by 
\begin{eqnarray}
  \label{eq:susyr}
  \delta X^{\Lambda} & = & \bar{\epsilon}^i \Omega_i^{\; \Lambda}\,,
  \,\nonumber\\ 
  \delta A_{\mu}{}^{\Lambda} & = & \varepsilon^{ij} \bar{\epsilon}_i
  \gamma_{\mu} \Omega_j{}^{\Lambda} + \varepsilon_{ij}
  \bar{\epsilon}^i \gamma_{\mu} \Omega^{j\, \Lambda}\,,\nonumber\\ 
  \delta \Omega_i{}^{\Lambda} & = & 2 \slash{\partial} 
  X^{\Lambda} \epsilon_i + \ft12 \gamma^{\mu \nu}
  F^-_{\mu\nu}{}^\Lambda \varepsilon_{ij} \epsilon^j +
  Y_{ij}{}^{\Lambda} \epsilon^j\,, \nonumber\\
  \delta Y_{ij}{}^{\Lambda} & = & 2 \bar{\epsilon}_{(i}
  \slash{\partial}\Omega_{j)}{}^{\Lambda} + 2 \varepsilon_{ik} 
  \varepsilon_{jl}\, \bar{\epsilon}^{(k} \slash{\partial}\Omega^{l)
  \Lambda}   \,. 
\end{eqnarray}

In the absence of charged fields, abelian gauge fields
$A_\mu{}^\Lambda$ appear exclusively through the field strengths,
${F}_{\mu\nu}{}^\Lambda = 2\,\partial_{[\mu}A_{\nu]}{}^\Lambda$ (we
consider Lagrangians that are at most quadratic in derivatives). The
field equations for these fields and the Bianchi identities for the
field strengths comprise $2n$ equations,
\begin{equation}
  \label{eq:eom-bianchi}
  \partial_{[\mu} {F}_{\nu\rho]}{}^\Lambda  
   = 0 = \partial_{[\mu} {G}_{\nu\rho]\,\Lambda} \,, 
\end{equation}
where 
\begin{equation}
  \label{eq:def-G}
  {G}_{\mu\nu\,\Lambda} =  
  \varepsilon_{\mu\nu\rho\sigma}\,
  \frac{\partial \mathcal{L}}{\partial{F}_{\rho\sigma}{}^\Lambda}
  \;.
\end{equation}
In the case at hand this implies,
\begin{equation}
  \label{eq:G-explicit}
  G^-_{\mu\nu \Lambda} = F_{\Lambda\Sigma} \,F^-_{\mu\nu}{}^\Sigma -
  \ft18 F_{\Lambda\Sigma\Gamma} \, \bar \Omega_i{}^{\Sigma}
  \gamma_{\mu\nu} \Omega_j{}^{\Gamma}\,\varepsilon^{ij} \,. 
\end{equation}
It is convenient to combine the tensors $F_{\mu\nu}{}^\Lambda$ and
$G_{\mu\nu\Lambda}$ into a $2n$-dimensional vector, 
\begin{equation}
  \label{eq:GM}
  G_{\mu\nu}{}^M= \pmatrix{{F}_{\mu\nu}{}^\Lambda\cr   
    \noalign{\vskip 1.5mm}
  {G}_{\mu\nu\Lambda} } \,, 
\end{equation}
so that (\ref{eq:eom-bianchi}) reads $\partial_{[\mu}
{G}_{\nu\rho]}{}^M = 0$. Obviously these $2n$ equations are invariant
under real $2n$-dimensional rotations of the tensors $G_{\mu\nu}{}^M$, 
\begin{equation}
  \label{eq:em-duality}
  \pmatrix{{F}^\Lambda\cr   \noalign{\vskip 1.5mm} {G}_\Lambda}
  \longrightarrow  
  \pmatrix{U^\Lambda{}_\Sigma & Z^{\Lambda\Sigma} \cr
    \noalign{\vskip 1.5mm} 
    W_{\Lambda\Sigma} & V_\Lambda{}^\Sigma }   
  \pmatrix{{F}^\Sigma\cr \noalign{\vskip 1.5mm}  {G}_\Sigma} \,.
\end{equation}
Half of the rotated tensors can be adopted as new field strengths
defined in terms of new gauge fields, and the Bianchi identities on
the remaining tensors can then be interpreted as field equations
belonging to some new Lagrangian expressed in terms of the new field
strengths. In order that such a Lagrangian exists, the real matrix in
(\ref{eq:em-duality}) must belong to the group ${\rm
  Sp}(2n;\mathbb{R})$. This group consists of real matrices that leave
the skew-symmetric tensor $\Omega_{MN}$ invariant,
\begin{equation}
  \label{eq:omega}
  \Omega = \left( \begin{array}{cc} 
0 & {\bf 1}\\ \!-{\bf 1} & 0 
\end{array} \right)  \;.  
\end{equation}
The conjugate matrix $\Omega^{MN}$ is defined by
$\Omega^{MN}\Omega_{NP}= - \delta^M{}_P$. Here we employ an
$\mathrm{Sp}(2n,\mathbb{R})$ covariant notation for the
$2n$-dimensional symplectic indices $M,N,\ldots$, such that $Z^M=
(Z^\Lambda, Z_\Lambda)$. Likewise we use vectors with lower indices
according to $Y_M= (Y_\Lambda,Y^\Lambda)$, transforming according to
the conjugate representation so that $Z^M\,Y_M$ is invariant. 

The ${\rm Sp}(2n;\mathbb{R})$ transformations are known as
electric/magnetic dualities, which also act on electric and magnetic
charges (for a review of electric/magnetic duality, see
\cite{deWit:2001pz}). The Lagrangian depends on the electric/magnetic
duality frame and is therefore not unique.\footnote{
  Up to terms proportional to the field equations of the vector fields
  and the auxiliary fields, the Lagrangian is covariant under
  electric/magnetic duality. } 
Different Lagrangians related by electric/magnetic duality lead to
equivalent field equations and thus belong to the same equivalence
class. These alternative Lagrangians remain supersymmetric and when
applying suitable redefinitions to the other fields, they can again be
brought into the form (\ref{eq:vlagr-kin}), characterized by a new
holomorphic function $F(X)$. In other words, different functions
$F(X)$ can belong to the same equivalence class. The new function is
such that the vector $X^M=(X^\Lambda,F_\Lambda)$ transforms under
electric/magnetic duality according to
\begin{equation}
  \label{eq:em-duality-X}
  \pmatrix{{X}^\Lambda\cr   \noalign{\vskip 1.5mm} {F}_\Lambda}
  \longrightarrow  
  \pmatrix{\tilde{X}^\Lambda\cr \noalign{\vskip 1.5mm} 
    {\tilde F}_\Lambda} = 
  \pmatrix{U^\Lambda{}_\Sigma & Z^{\Lambda\Sigma} \cr
    \noalign{\vskip 1.5mm}
    W_{\Lambda\Sigma} & V_\Lambda{}^\Sigma }   
  \pmatrix{{X}^\Sigma\cr \noalign{\vskip 1.5mm} {F}_\Sigma} \,.
\end{equation}
The new function $\tilde F(\tilde X)$ of the new scalars $\tilde
X^\Lambda$ follows from integration of (\ref{eq:em-duality-X}) and
takes the form
\begin{eqnarray}
  \label{eq:new-F}
  \tilde F(\tilde X)&=& F(X) -\ft12 X^\Lambda F_\Lambda(X) 
  + \ft12 (U^\mathrm{T} W)_{\Lambda\Sigma} X^\Lambda X^\Sigma  \nonumber
  \\
  &&
  + \ft12 (U^\mathrm{T} V+ W^\mathrm{T} Z)_\Lambda{}^\Sigma  X^\Lambda
  F_\Sigma(X)  
  + \ft12 (Z^\mathrm{T} V)^{\Lambda\Sigma} F_\Lambda(X) F_\Sigma(X)\,,
\end{eqnarray}
up to a constant and to terms linear in the $\tilde X^\Lambda$. These
terms, which will be ignored in what follows, cannot be present in the
case of local supersymmetry. In general it is not easy to determine
$\tilde F(\tilde X)$ from (\ref{eq:new-F}) as it involves the
inversion of $\tilde X^\Lambda = U^\Lambda{}_\Sigma X^\Sigma +
Z^{\Lambda\Sigma} F_\Sigma(X)$. The duality transformations on higher
derivatives of $F(X)$ follow by differentiation and we note the
results \cite{deWit:1996ix},
\begin{eqnarray}
  \label{eq:dual-higher-F-der}
  \tilde F_{\Lambda\Sigma}(\tilde X) &=& (V_\Lambda{}^\Gamma
  F_{\Gamma\Xi} + W_{\Lambda\Xi} )\, [\mathcal{S}^{-1}]^\Xi{}_\Sigma\,,
  \nonumber\\ 
  \tilde F_{\Lambda\Sigma\Gamma}(\tilde X) &=& F_{\Xi\Delta\Omega}\,
  [\mathcal{S}^{-1}]^\Xi{}_\Lambda\,[\mathcal{S}^{-1}]^\Delta{}_\Sigma\,
  [\mathcal{S}^{-1}]^\Omega{}_\Gamma\,, 
\end{eqnarray}
where $\mathcal{S}^\Lambda{}_\Sigma = \partial \tilde
X^\Lambda/\partial X^\Sigma = U^\Lambda{}_\Sigma + Z^{\Lambda\Gamma}
F_{\Gamma\Sigma}$. From the first equation one derives, 
\begin{equation}
  \label{eq:dual-N}
  \tilde N_{\Lambda\Sigma}(\tilde X,\tilde{\bar X}) =
  N_{\Gamma\Delta}\, [\mathcal{S}^{-1}]^\Gamma{}_\Lambda\,
  [\bar\mathcal{S}^{-1}]^\Delta{}_\Sigma \,. 
\end{equation}

To determine the action of the dualities on the fermions, we consider
supersymmetry transformations of $X^M= (X^\Lambda, F_\Lambda)$, which
take the form $\delta X^M = \bar\epsilon^i \Omega_i{}^M$, thus
defining an $\mathrm{Sp}(2n,\mathbb{R})$ covariant fermionic vector
$\Omega_i{}^M$, 
\begin{equation}
  \label{eq:symplectic-fermion}
  \Omega_i{}^M = \pmatrix{ \Omega_i{}^\Lambda \cr
    \noalign{\vskip 1.5mm}
  F_{\Lambda\Sigma}\, \Omega_i{}^\Sigma}  \;.
\end{equation}
Complex conjugation leads to a second vector, $\Omega^i{}^M$, of
opposite chirality. From (\ref{eq:symplectic-fermion}) one derives
directly that, under electric/magnetic duality,
\begin{equation}
  \label{eq:em-fermion}
  \tilde\Omega_i{}^\Lambda =
  \mathcal{S}^\Lambda{}_\Sigma\,\Omega_i{}^\Sigma\,. 
\end{equation}
With this result one can show that (\ref{eq:em-duality}),
(\ref{eq:dual-higher-F-der}) and (\ref{eq:em-fermion}) are
consistent. 

The supersymmetry transformation of $\Omega_i{}^M$ takes the following
form,
\begin{equation}
  \label{eq:delta-Omega-M}
  \delta\Omega_i{}^M = 2\slash{\partial}X^M\epsilon_i + \ft12
  \gamma^{\mu\nu}G^-_{\mu\nu}{}^M\varepsilon_{ij}\epsilon^j +
  Z_{ij}{}^M \epsilon^j \,,
\end{equation}
where 
\begin{equation}
  \label{eq:symplectic-Z}
  Z_{ij}{}^M = \pmatrix{ Y_{ij}{}^\Lambda \cr
    \noalign{\vskip 1.5mm}
  F_{\Lambda\Sigma}\, Y_{ij}{}^\Sigma -\ft12 F_{\Lambda\Sigma\Gamma}
    \,\bar\Omega_i{}^\Sigma\Omega_j{}^\Gamma  }  \;.
\end{equation}
This suggests that $Z_{ij}{}^M$ transforms under electric/magnetic
duality as a symplectic vector. However, this is only possible
provided we drop the pseudo-reality constraint on $Y_{ij}{}^\Lambda$.
In that case imposing a pseudo-reality condition on $Z_{ij}{}^M$ is
manifestly consistent with ${\rm Sp}(2n;\mathbb{R})$ and implies both
the pseudo-reality of and the field equations associated with the
$Y_{ij}{}^\Lambda$.

The electric/magnetic duality transformations thus define equivalence
classes of Lagrangians. A subgroup thereof may constitute an
invariance of the theory \cite{Gaillard:1981rj}, meaning that the
Lagrangian and its underlying function $F(X)$ do not change
\cite{deWit:1984pk,Cecotti:1988qn}. More specifically, an invariance implies
\begin{equation}
  \label{eq:invariant-F}
  \tilde F(\tilde X)= F(\tilde X)\,,
\end{equation}
so that the result of the duality leads to a Lagrangian based on
$\tilde F(\tilde X)$ which is identical to the original Lagrangian.
Because $\tilde F(\tilde X)\not = F(X)$, as is obvious from
(\ref{eq:new-F}), $F(X)$ is not an invariant {\it function}. Instead
the above equation implies that the substitution $X^\Lambda\to \tilde
X^\Lambda$ into the function $F(X)$ and its derivatives, induces
precisely the duality transformations. For example, we obtain,
\begin{eqnarray}
  \label{eq:dual-symm-F-der}
  F_\Lambda (\tilde X) &=&
  V_\Lambda{}^\Sigma F_\Sigma(X) + W_{\Lambda\Sigma} X^\Sigma \,,
  \nonumber\\ 
  F_{\Lambda\Sigma}(\tilde X) &=& (V_\Lambda{}^\Gamma
  F_{\Gamma\Xi} + W_{\Lambda\Xi} )\, [\mathcal{S}^{-1}]^\Xi{}_\Sigma\,,
  \nonumber\\ 
  F_{\Lambda\Sigma\Gamma}(\tilde X) &=& F_{\Xi\Delta\Omega}\,
  [\mathcal{S}^{-1}]^\Xi{}_\Lambda\,[\mathcal{S}^{-1}]^\Delta{}_\Sigma\,
  [\mathcal{S}^{-1}]^\Omega{}_\Gamma\,. 
\end{eqnarray}

We elucidate these invariances for the subgroup that acts linearly on
the gauge fields $A_\mu{}^\Lambda$. These symmetries are characterized
by the fact that the matrix in (\ref{eq:em-duality}) and
(\ref{eq:em-duality-X}) has a block-triangular form with $V=
[U^{\mathrm{T}}]^{-1}$ and $Z=0$. Hence this is not a general duality
as the Lagrangian is still based on the same gauge fields, up to the
linear transformation $A_\mu{}^\Lambda\to \tilde A_\mu{}^\Lambda=
U^\Lambda{}_\Sigma A_\mu{}^\Sigma$. Note that all fields in the Lagrangian
(\ref{eq:vlagr-kin}) carry upper indices and are thus subject to the
same linear transformation. The function $F(X)$ changes with an
additive term which is a quadratic polynomial with real coefficients.
\begin{equation}
  \label{eq:F-PQ}
  \tilde F(\tilde X) = F (U^\Lambda{}_\Sigma X^\Sigma) = F(X) + \ft12
  (U^\mathrm{T}   W)_{\Lambda\Sigma}\, X^\Lambda X^\Sigma \,.  
\end{equation}
This term induces a total derivative term in the Lagrangian, equal to
\begin{equation}
  \label{eq:variationPQ}
  \mathcal{L} \to  \mathcal{L} -\ft18 \mathrm{i} 
  \varepsilon^{\mu\nu\rho\sigma} 
  (U^\mathrm{T} W)_{\Lambda\Sigma}\,
  F_{\mu\nu}{}^\Lambda F_{\rho\sigma}{}^\Sigma\,.  
\end{equation}

\subsection{Gauge transformations}
\label{sec:gauge-transformations}
Non-abelian gauge groups will act non-trivially on the vector fields
and must therefore involve a subgroup of the duality group.  The
electric gauge fields $A_\mu{}^\Lambda$ associated with this gauge
group are provided by vector multiplets. Because the duality group
acts on both electric and magnetic charges, in view of the fact that
it mixes field strengths with dual field strengths as shown by
(\ref{eq:em-duality}), we will eventually introduce magnetic gauge
fields $A_{\mu \Lambda}$ as well, following the procedure explained in
\cite{deWit:2005ub}. The $2n$ gauge fields $A_\mu{}^M$ will then
comprise both type of fields, $A_\mu{}^M= (A_\mu{}^\Lambda,
A_{\mu\Lambda})$. The role played by the magnetic gauge fields will be
clarified later. For the moment one may associate $A_{\mu\,\Lambda}$
with the dual field strengths ${G}_{\mu\nu \,\Lambda}$, by writing
${G}_{\mu\nu\,\Lambda} \equiv 2 \,\partial_{[\mu} A_{\nu]\Lambda}$.

The generators (as far as their embedding in the duality group is
concerned) are defined as follows. The generators of the subgroup that
is gauged, are $2n$-by-$2n$ matrices $T_M$, where we are assuming the
presence of both electric and magnetic gauge fields, so that the
generators decompose according to $T_M=(T_{\Lambda},T^\Lambda)$.
Obviously $T_{\Lambda N}{}^P$ and $T^\Lambda{}_N{}^P$ can be
decomposed into the generators of the duality group and are thus of
the form specified in (\ref{eq:em-duality}). Denoting the gauge group
parameters by $\Lambda^M(x) = (\Lambda^\Lambda(x),
\Lambda_\Lambda(x))$, $2n$-dimensional ${\rm Sp}(2n;\mathbb{R})$
vectors $Y^M$ and $Z_M$ transform according to
\begin{equation}
  \label{eq:gauge-tr-Y-Z}
  \delta Y^M = -g \Lambda^N \,T_{NP}{}^M \,Y^P\,,\qquad
    \delta Z_M = g \Lambda^N \,T_{NM}{}^P \,Z_P\,, 
\end{equation}
where $g$ denotes a universal gauge coupling constant. Covariant
derivatives thus take the form,
\begin{eqnarray}
  \label{eq:cov-derivative}
  D_\mu Y^M &=& \partial_\mu Y^M + g A_\mu{}^N\, T_{NP}{}^M\,Y^P
  \nonumber \\ 
  &=& \partial_\mu Y^M + g A_\mu{}^\Lambda\, T_{\Lambda P}{}^M\,Y^P +
  g A_{\mu\Lambda}\, T^\Lambda{}_{P}{}^M\,Y^P \,, 
\end{eqnarray}
and similarly for $D_\mu Z_M$. The gauge fields then transform
according to 
\begin{equation}
  \label{eq:gauge-tr-A}
  \delta A_\mu{}^M = \partial_\mu \Lambda^M + g\, T_{PQ}{}^M
  A_\mu{}^P\, \Lambda^Q \,.   
\end{equation}

For clarity we first consider electric gaugings where the gauge
transformations have a block-triangular form and there are only
electric gauge fields. Hence we ignore the fields $A_{\mu\Lambda}$ and
assume $T^\Lambda{}_N{}^P=0$ and $T_\Lambda{}^{\Sigma\Gamma}= 0$. All
the fields in the Lagrangian carry upper indices, so that they will
transform as in $\delta X^\Lambda = - g \Lambda^\Gamma
T_{\Gamma\Sigma}{}^\Lambda \,X^\Sigma$.  The transformation rule for
$A_\mu{}^\Lambda$ given above is in accord with this expression,
provided we assume that $T_{\Gamma\Sigma}{}^\Lambda$ is antisymmetric
in $\Gamma$ and $\Sigma$. This has to be the case here as consistency
requires that the $T_{\Gamma\Sigma}{}^\Lambda$ are structure constants
of the non-abelian group.  In the more general situation discussed in
later sections, this is not necessarily the case. The embedding into
$\mathrm{Sp}(2n,\mathbb{R})$ implies furthermore that
$T_{\Lambda\Sigma}{}^\Gamma = - T_\Lambda{}^\Sigma{}_\Gamma$, while
the nonvanishing left-lower block $T_{\Lambda\Sigma\Gamma}$ is
symmetric in $\Sigma$ and $\Gamma$.

Furthermore we note that (\ref{eq:F-PQ}) implies 
\begin{equation}
  \label{eq:delta-F}
  F_\Lambda(X)\,\delta X^\Lambda = -g \Lambda^\Gamma
  T_{\Gamma\Sigma}{}^\Lambda\, 
  F_\Lambda(X)\,X^\Sigma = - \ft12g\,\Lambda^\Lambda
\,T_{\Lambda\Sigma\Gamma} X^\Sigma X^\Gamma \,. 
\end{equation}
Upon replacing $\Lambda^\Lambda$ with $X^\Lambda$ we conclude that
the fully symmetric part of $T_{\Lambda\Sigma\Gamma}$ vanishes. This,
and the closure of the gauge group, leads to the following three
equations,
\begin{eqnarray}
  \label{eq:T-triangle}
  &&
  T_{(\Lambda\Sigma\Gamma)}= 0\,, \nonumber\\
  &&
  T_{[\Lambda\Sigma}{}^\Delta\,T_{\Gamma]\Delta}{}^\Xi = 0 \,,
  \nonumber\\ 
  &&
  4\,T_{(\Gamma[\Lambda}{}^\Delta \, T_{\Sigma]\Xi)\Delta} 
  - T_{\Lambda\Sigma}{}^\Delta T_{\Delta\Gamma\Xi} = 0 \, .
\end{eqnarray}
The variation of the Lagrangian (\ref{eq:variationPQ}) under gauge
transformations now takes the form 
\begin{equation}
  \label{eq:gauge-variationPQ}
  \mathcal{L} \to  \mathcal{L} + \ft18 \mathrm{i} \, 
  \varepsilon^{\mu\nu\rho\sigma}  \,\Lambda^\Lambda\, 
  T_{\Lambda\Sigma\Gamma}\,
  \mathcal{F}_{\mu\nu}{}^\Sigma \mathcal{F}_{\rho\sigma}{}^\Gamma\,,
\end{equation}
where the tensors $\mathcal{F}_{\mu\nu}{}^\Lambda$ denote the
non-abelian field strengths,
\begin{equation}
  \label{eq:nonabelian-FS}
  \mathcal{F}_{\mu\nu}{}^\Lambda = \partial_\mu A_\nu{}^\Lambda -
  \partial_\nu A_\mu{}^\Lambda  + g\, T_{\Sigma\Gamma}{}^\Lambda\,
  A_\mu{}^\Sigma A_\nu{}^\Gamma \,. 
\end{equation}
This result implies that (\ref{eq:gauge-variationPQ}) no longer
constitutes a total derivative in view of the space-time dependent
transformation parameters $\Lambda^\Lambda(x)$. Therefore its
cancellation requires to add a new type of term \cite{deWit:1984px},
\begin{equation}
  \label{eq:cs-electric}
  \mathcal{L} =  \ft13 \mathrm{i} g\, 
  \varepsilon^{\mu\nu\rho\sigma}\,T_{\Lambda\Sigma\Gamma} 
  \,A_\mu{}^\Lambda A_\nu{}^\Sigma (\partial_\rho A_\sigma{}^\Gamma 
  + \ft38 g\, T_{\Xi\Delta}{}^\Gamma \,A_\rho{}^\Xi A_\sigma{}^\Delta)
  \,. 
\end{equation}
No other terms in the action will depend on $T_{\Lambda\Sigma\Gamma}$.
At this point we should remind the reader that the gauging breaks
supersymmetry, unless one adds the standard masslike and potential
terms to the Lagrangian (\ref{eq:L-vector}), which involve the
$T_{\Lambda\Sigma}{}^\Gamma$. We present them below for completeness,
\begin{eqnarray}
  \label{eq:electric-masslike-potential}
  \mathcal{L}_g &=& - \ft12  g \,N_{\Lambda\Sigma}
  T_{\Gamma\Xi}{}^\Sigma  \Big[ \varepsilon^{ij}\, \bar\Omega_i{}^\Lambda 
  \Omega_j{}^\Gamma \bar X^\Xi + 
  \varepsilon_{ij} \, \bar\Omega^{i\Lambda} \Omega^{j\Gamma} X^\Xi
  \Big] \,,   \nonumber \\[1mm]
  \mathcal{L}_{g^2} &=& g^2\, N_{\Lambda\Sigma}\,
  T_{\Gamma\Xi}{}^\Lambda \bar X^\Gamma   X^\Xi \,
  T_{\Delta\Omega}{}^\Sigma \bar X^\Delta  X^\Omega\,.  
\end{eqnarray}
In later sections we will exhibit the generalization of these terms to
the case where both electric and magnetic charges are present. 
\subsection{Electric and magnetic charges}
\label{e+m-charges}
We now consider more general gauge groups without restricting
ourselves to electric charges. Therefore we include both electric
gauge fields $A_\mu{}^\Lambda$ and magnetic gauge fields
$A_{\mu\,\Lambda}$.  Only a subset of these fields is usually involved
in the gauging, but the additional magnetic gauge fields could
conceivably lead to new propagating degrees of freedom. We will
discuss in due course how this is avoided. In this subsection we
consider the scalar and spinor fields. The treatment of the vector
fields is more involved and is explained in section
\ref{sec:embedding-tensor}.

The charges $T_{MN}{}^P$ correspond to a more general subgroup of the
duality group. Hence they must take values in the Lie algebra
associated with $\mathrm{Sp}(2n,\mathbb{R})$, which implies,
\begin{equation}
  \label{eq:sp-constraint}
  T_{M[N}{}^Q\,\Omega_{P]Q} =0\,.
\end{equation}
Combining the two equations (\ref{eq:new-F}) and
(\ref{eq:invariant-F}) leads to the condition \cite{deWit:1984pk},
\begin{equation}
  \label{eq:symplectic-invariance} 
  T_{MN}{}^Q \Omega_{PQ} \,X^NX^P = 
  T_{M\Lambda\Sigma} X^\Lambda X^\Sigma -2 T_{M\Lambda}{}^\Sigma X^\Lambda
  F_\Sigma - T_M{}^{\Lambda\Sigma}F_\Lambda F_\Sigma =0\,.  
\end{equation} 
This result can also be written as 
\begin{equation}
  \label{eq:dlta-F}
  F_\Lambda \delta X^\Lambda = -\ft12 \Lambda^M \Big( T_{M\Lambda\Sigma}
  X^\Lambda X^\Sigma 
  +T_M{}^{\Lambda\Sigma} F_\Lambda F_\Sigma\Big) \,, 
\end{equation}
which generalizes (\ref{eq:delta-F}). Furthermore we impose the
so-called representation constraint \cite{deWit:2005ub}, which implies that we
suppress a representation of the rigid symmetry group in $T_{MN}{}^P$,
\begin{equation}
  \label{eq:lin}
  T_{(MN}{}^{Q}\,\Omega_{P)Q} =0 
\Longrightarrow  \left\{
\begin{array}{l}
T^{(\Lambda\Sigma\Gamma)}=0\,,\\[.2ex]
2T^{(\Gamma\Lambda)}{}_{\Sigma}= 
T_{\Sigma}{}^{\Lambda\Gamma}\,, \\[.2ex]
T_{(\Lambda\Sigma\Gamma)}=0\,,\\[.2ex]
2T_{(\Gamma\Lambda)}{}^{\Sigma}=
T^{\Sigma}{}_{\Lambda\Gamma}\,.
\end{array}
\right.
\end{equation}
This constraint is a generalization of the first equation
(\ref{eq:T-triangle}). Observe that the generators
$T_{\Lambda\Sigma}{}^\Gamma$ are no longer antisymmetric in $\Lambda$
and $\Sigma$, a feature that we will discuss in more detail in section
\ref{sec:embedding-tensor}.

The action of electric/magnetic duality on the fermions was already
discussed earlier when introducing the $\mathrm{Sp}(2n,\mathbb{R})$
covariant fermionic vector $\Omega_i{}^M$ (c.f.
(\ref{eq:symplectic-fermion})). In terms of this field we can rewrite
the Lagrangian (\ref{eq:L-matter}) in a compact form,
\begin{equation}
  \label{eq:lagrangian-pieces}
  \mathcal{L}_{\mathrm{matter}} = -\mathrm{i} \Omega_{MN}\,
  \partial_\mu X^M \,\partial^\mu \bar   X^N + 
  \ft14\mathrm{i}\Omega_{MN}\Big[\bar
  \Omega^{iM} \slash{\partial} \Omega_i{}^N
  -\bar \Omega_i{}^M \slash{\partial} \Omega^{iN} \Big] \,.  
\end{equation} 
In the expressions on the right-hand side it is straightforward to
 replace the ordinary derivatives by the covariant ones defined in
 (\ref{eq:cov-derivative}), i.e., 
\begin{eqnarray}
  \label{eq:covariant-symplectic}
  D_\mu X^M&=&\partial_\mu X^M +g\, A_\mu{}^N \,T_{NP}{}^M \,X^P\,,
  \nonumber \\ 
  D_\mu \Omega_i{}^M &=&\partial_\mu\Omega_i{}^M +g\, A_\mu{}^N
  \,T_{NP}{}^M \,\Omega_i{}^P \,,  
\end{eqnarray}
and evaluate the gauge couplings. In particular we can then compare to
the results of subsection \ref{sec:gauge-transformations}, where we
considered only electric gauge fields with charges restricted by
$T_{\Lambda}{}^{\Sigma\Gamma}=0$. To do this systematically we note
the identity, 
\begin{equation}
  \label{eq:derivative-invariance}
  T_{MN\Lambda} X^N - F_{\Lambda\Sigma} \,T_{MN}{}^\Sigma X^N =0\,. 
\end{equation}
This equation can also be written as $F_{\Lambda\Sigma}\, \delta
X^\Sigma = - g \Lambda^M T_{MN\Lambda} X^N$, which is the infinitesimal
form of the first equation (\ref{eq:dual-symm-F-der}).
Alternatively it can be derived from (\ref{eq:symplectic-invariance})
upon differentiation with respect to $X^\Lambda$. 

It is possible to cast (\ref{eq:derivative-invariance}) in a
symplectically covariant form by introducing a vector $U^M =
(U^\Lambda, F_{\Sigma\Gamma} U^\Gamma)$, so that
\begin{equation}
  \label{eq:cov-der-invariance}
  \Omega_{MQ} T_{NP}{}^Q \,X^P \,U^M=0\,,
\end{equation}
for any such vector $U^M$. This form is convenient in calculations
presented later.

From (\ref{eq:derivative-invariance}) one easily derives that $D_\mu
X_\Lambda = D_\mu F_\Lambda = F_{\Lambda\Sigma} \, D_\mu X^\Sigma$,
which enables one to derive
\begin{equation}
  \label{eq:scalar-kinetic}
  -\mathrm{i} \Omega_{MN}\, D_\mu X^M \,D^\mu \bar
  X^N = -N_{\Lambda\Sigma} \, D_\mu X^\Lambda \, D^\mu \bar
  X^\Sigma\,. 
\end{equation}
This result shows that the generators $T_{M\Lambda\Sigma}$ are absent,
in accord with what was found in subsection
\ref{sec:gauge-transformations}. 

Next we consider the gauge field interactions with the fermions. It is
convenient to first derive an additional identity, which follows from
taking a supersymmetry variation of (\ref{eq:derivative-invariance}), 
\begin{equation}
  \label{eq:fermion-der-invariance}
  T_{MN\Lambda} \Omega_i{}^N = F_{\Lambda\Sigma} \,T_{MN}{}^\Sigma
    \Omega_i{}^N +F_{\Lambda\Sigma\Gamma}\,\Omega_i{}^\Sigma
    \,T_{MN}{}^\Gamma X^N   \,.  
\end{equation}
This result can be obtained from the infinitesimal form of the third
equation of (\ref{eq:dual-symm-F-der}).
Using this equation one verifies that $D_\mu \Omega_{i\Lambda} =
F_{\Lambda\Sigma} \,D_\mu \Omega_i{}^\Sigma + F_{\Lambda\Sigma\Gamma}\,
\Omega_i{}^\Gamma D_\mu X^\Sigma$, which leads to 
\begin{eqnarray}
  \label{eq:fermion-kinetic}
    \ft14\mathrm{i}\Omega_{MN}\Big[\bar
    \Omega^{iM} \Slash{D} \Omega_i{}^N
    -\bar \Omega_i{}^M \Slash{D} \Omega^{iN} \Big]  &\!=\!& -\ft14
    N_{\Lambda\Sigma}\Big(\bar\Omega^{i\Lambda}\Slash{D}
    \Omega_i{}^\Sigma 
      +\bar \Omega_i{}^\Lambda \Slash{D}\Omega ^{i\Sigma}\Big)
      \nonumber\\
      &&
      -\ft14 \mathrm{i}\Big( F_{\Lambda\Sigma\Gamma}\bar
    \Omega_i{}^\Lambda\Slash{D} X^\Sigma \Omega ^{i\Gamma} 
      - \bar F_{\Lambda\Sigma\Gamma} \bar\Omega^{i
    \Lambda} \Slash{D} \bar X^\Sigma \Omega _i{}^\Gamma\Big) \,. 
\end{eqnarray}
Again the generator $T_{M\Lambda\Sigma}$ is absent in the expression
above. The results of this subsection explain how to introduce the
electric and magnetic charges, but in no way ensure the gauge
invariance or the supersymmetry of the Lagrangian. To obtain such a
result we first need to explain some more general features of theories
with both electric and magnetic gauge fields in four space-time
dimensions. This is the topic of the following section.

As a side remark we note that the Killing potential (or moment map)
associated with the isometries considered above, takes the form,
\begin{equation}
  \label{eq:U(1)-moment-map}
  \nu_M = T_{MN}{}^Q \Omega_{PQ} \bar X^N X^P \,.
\end{equation}
Indeed, making use again of (\ref{eq:derivative-invariance}), one
straightforwardly derives $\partial_\Lambda \nu_M = \mathrm{i}
N_{\Lambda\Sigma} \,\delta\bar X^\Sigma$.

Finally we return to the gauge transformations of the auxiliary fields
$Y_{ij}{}^\Lambda$, which can be derived by requiring that the
Lagrangian (\ref{eq:vlagr-Y}) is gauge invariant. A straightforward
calculation lead to the following result, 
\begin{equation}
  \label{eq:gauge-Y}
  \delta Y_{ij}{}^\Lambda= - \ft12 \Lambda^M T_{MN}{}^\Lambda (
  Z_{ij}{}^N+ \varepsilon_{ik}\varepsilon_{jl} \,Z^{klN}) \,, 
\end{equation}
where $Z_{ij}{}^M$ was defined in (\ref{eq:symplectic-Z}). Note that
this result is in accord with the electric/magnetic dualities
suggested for $Z_{ij}{}^M$.

\section{The gauge group and the embedding tensor}
\label{sec:embedding-tensor} 
\setcounter{equation}{0} 
Here we follow \cite{deWit:2005ub} and discuss the embedding of possible gauge
groups into the rigid invariance group $\mathrm{G}_{\mathrm{rigid}}$
of the theory. In the context of this paper, the latter is often a
product group as the vector multiplets and the hypermultiplets are
invariant under independent symmetry groups. As explained in the
previous section the non-abelian gauge transformations on the vector
multiplets must be embedded into the electric/magnetic duality
group.

It is convenient to discuss group embeddings in terms of a so-called
embedding tensor $\Theta_M{}^{\sf a}$ which 
specifies the decomposition of the gauge group generators $T_M$ into
the generators associated with the full rigid invariance group
 $\mathrm{G}_{\mathrm{rigid}}$, 
\begin{equation}
  \label{eq:T-into-t}
  T_M = \Theta_M{}^{\sf a} \,t_{\sf a} \,.
\end{equation}
Not all the gauge fields have to be involved in the gauging, so
generically the embedding tensor projects out certain combinations of
gauge fields; the rank of the tensor determines the dimension of the
gauge group, up to central extensions associated with abelian factors.
Decomposing the embedding tensor as $\Theta_M{}^{\sf a} =
(\Theta_\Lambda{}^{\sf a}, \Theta^{\Lambda\,{\sf a}})$, covariant
derivatives take the form,
\begin{equation}
  \label{eq:cov-der}
   D_\mu\equiv 
  \partial_{\mu}-g A_{\mu}{}^{M} T_M = \partial_{\mu}-g
  A_{\mu}{}^{\Lambda}\Theta_{\Lambda}{}^{{\sf a}} \,t_{{\sf a}} -g
  A_{\mu\,\Lambda}\Theta^{\Lambda\,{\sf a}} \,t_{{\sf a}} \;.  
\end{equation}
The embedding tensor will be regarded as a spurionic object which can
be assigned to a (not necessarily irreducible) representation of
the rigid invariance group $\mathrm{G}_{\mathrm{rigid}}$.

It is known that a number of ($\mathrm{G}_{\mathrm{rigid}}$-covariant)
constraints must be imposed on the embedding tensor. We already
encountered the representation constraint (\ref{eq:lin}), which is
linear in the embedding tensor. Two other constraints are quadratic in
the embedding tensor and read, 
\begin{eqnarray}
  f_{\sf ab}{}^{\sf c}\, \Theta_{M}{}^{\sf a}\,\Theta_{N}{}^{\sf b}
+(t_{{\sf a}})_{N}{}^{P}\,\Theta_{M}{}^{\sf a}\Theta_{P}{}^{\sf c} &=&0\,, 
  \label{eq:clos}  \\[1ex]
\Omega^{MN}\,\Theta_{M}{}^{\sf a}\Theta_{N}{}^{\sf b}~=~0
\;\;\Longleftrightarrow\; \;
\Theta^{\Lambda\,[\sf a}\Theta_{\Lambda}{}^{\sf b]} &=&0 \;,
\label{eq:quad}
\end{eqnarray}
where the $f_{{\sf a}{\sf b}}{}^{\sf c}$ are the structure constants
associated with the group $\mathrm{G}$.  The first constraint is
required by the closure of the gauge group generators. Indeed, from
(\ref{eq:clos}) it follows that the gauge algebra generators close
according to
\begin{equation}
  \label{eq:closure}
  {}[T_{M},T_{N}] = -T_{MN}{}^{P}\,T_{P} \;,
\end{equation}
where the structure constants of the gauge group coincide with
$T_{MN}{}^{P}\equiv \Theta_{M}{}^{{\sf a}} \,(t_{{\sf a}})_{N}{}^{P}$ up
to terms that vanish upon contraction with the embedding tensor
$\Theta_P{}^{\sf a}$.  We recall that the $T_{MN}{}^P$ generate a
subgroup of $\mathrm{Sp}(2n,\mathbb{R})$ in the $(2n)$-dimensional
representation, so that they are subject to the condition
(\ref{eq:sp-constraint}).  In electric/magnetic components the latter
condition corresponds to $T_{M\Lambda}{}^\Sigma=
-T_{M}{}^\Sigma{}_\Lambda$, $T_{M \Lambda\Sigma}=T_{M\Sigma\Lambda}$
and $T_{M}{}^{\Lambda\Sigma}=T_{M}{}^{\Sigma\Lambda}$.

Note that (\ref{eq:clos}) implies that the embedding tensor is gauge
invariant, while the second quadratic constraint (\ref{eq:quad})
implies that the charges are mutually local, so that an
electric/magnetic duality exists that converts all the charges to
electric ones.  These two quadratic constraints are not completely
independent, as can be seen from symmetrizing the constraint
(\ref{eq:clos}) in $(MN)$ and making use of the linear conditions
(\ref{eq:lin}) and (\ref{eq:sp-constraint}). This leads to
\begin{eqnarray}
  \label{eq:constraint-eq}
\Omega^{MN}\,\Theta_{M}{}^{{\sf a}}\Theta_{N}{}^{{\sf b}}\,
(t_{{\sf b}})_{P}{}^{Q}&=&0 \;. 
\end{eqnarray}
This shows that, for non-vanishing $(t_{{\sf b}})_{P}{}^{Q}$, the
second quadratic constraint~(\ref{eq:quad}) is in fact a consequence
of the other constraints. The constraint (\ref{eq:quad}) is only an
independent constraint when ${\sf a}$ and ${\sf b}$ do not refer to
generators that act on the vector multiplets. This issue is relevant
here as $\mathrm{G}_{\mathrm{rigid}}$ may contain independent generators
that act exclusively in the matter (i.e., hypermultiplet) sector.

A further consequence of~(\ref{eq:lin}) is the equation  
\begin{equation}
  \label{eq:Z-d}
  T_{(MN)}{}^{P}= Z^{P,{\sf a}}\, d_{{\sf a} \,MN} \;, 
\end{equation}
with
\begin{eqnarray}
  \label{eq:def-Z-d}
d_{{\sf a}\, MN} &\equiv& (t_{\sf a})_M{}^P\, \Omega_{NP}\,,\nonumber\\
Z^{M,{\sf a}}&\equiv&\ft12\Omega^{MN}\Theta_{N}{}^{{\sf a}}
\quad
\Longrightarrow
\quad
\left\{
\begin{array}{rcr}
Z^{\Lambda{\sf a}} &=& \ft12\Theta^{\Lambda{\sf a}} \,,\\[1ex]
Z_{\Lambda}{}^{{\sf a}} &=& -\ft12\Theta_{\Lambda}{}^{{\sf a}} \,,
\end{array}
\right.
\end{eqnarray}
so that $d_{{\sf a}\,MN}$ defines a
$\mathrm{G}_{\mathrm{rigid}}$-invariant tensor symmetric in $(MN)$.
The gauge invariant tensor $Z^{M,{\sf a}}$ will serve as a projector on
the tensor fields to be introduced below \cite{deWit:2005hv}. We note
that the constraint (\ref{eq:quad}) can now be written as, 
\begin{equation}
  \label{eq:Z-Theta}
  Z^{M,{\sf a}} \,\Theta_M{}^{\sf b} =0\,.
\end{equation}

Let us return to the closure relation (\ref{eq:closure}). Although
the left-hand side is antisymmetric in $M$ and $N$, this does not
imply that $T_{MN}{}^P$ is antisymmetric as well, but only that its symmetric
part vanishes upon contraction with the embedding tensor. Indeed, this is
reflected by (\ref{eq:Z-d}) and (\ref{eq:Z-Theta}). Consequently,
the Jacobi identity holds only modulo terms that vanish upon
contraction with the embedding tensor, as is shown explicitly by 
\begin{equation}
  \label{Jacobi-X}
  {T_{[MN]}{}^P\, T_{[QP]}{}^R + T_{[QM]}{}^P\, T_{[NP]}{}^R  +
 T_{[NQ]}{}^P \, T_{[MP]}{}^R} = - Z^{R,{\sf a}}\,d_{{\sf a}\,P[Q}\,
 T_{MN]}{}^P \,.   
\end{equation}
To compensate for this lack of closure and, at the same time, to avoid
unwanted degrees of freedom, we introduce an extra gauge invariance
for the gauge fields, in addition to the usual nonabelian gauge
transformations, 
\begin{equation}
  \label{eq:A-var}
  \delta A_\mu{}^M =  D_\mu\Lambda^M-
  g\,Z^{M,{\sf a}}\,\Xi_{\mu\,{\sf a}} \,, 
\end{equation}
where the $\Lambda^M$ are the gauge transformation parameters and the
covariant derivative reads, $D_\mu\Lambda^M =\partial_\mu\Lambda^M +
g\, T_{PQ}{}^M\,A_\mu{}^P\Lambda^Q$. The transformations proportional
to $\Xi_{\mu\,{\sf a}}$ enable one to gauge away those vector fields
that are in the sector of the gauge generators $T_{MN}{}^P$ where the
Jacobi identity is not satisfied (this sector is perpendicular to the
embedding tensor by virtue of (\ref{eq:Z-Theta})). Note that the
covariant derivative is invariant under the transformations
parametrized by $\Xi_{\mu\,{\sf a}}$, because of the contraction of the
gauge fields $A_\mu{}^M$ with the generators $T_M$. The gauge
symmetries parametrized by the functions $\Lambda^M(x)$ and
$\Xi_{{\sf a}\mu}(x)$ form a group, as follows from the commutation
relations,
\begin{eqnarray}
  \label{eq:gauge-commutators}
  {}[\delta(\Lambda_1),\delta(\Lambda_2)] &=& \delta(\Lambda_3) +
  \delta(\Xi_3) \,,  \nonumber \\
  {}[\delta(\Lambda),\delta(\Xi)] &=& \delta(\tilde\Xi) \,, 
\end{eqnarray}
where 
\begin{eqnarray}
  \label{eq:gauge-parameters}
  \Lambda_3{}^M &=& g\,T_{[NP]}{}^M \Lambda_2^N\Lambda_1^P\,, \nonumber\\
  \Xi_{3 \mu{\sf a}} &=& d_{{\sf a} NP}( \Lambda_1^N
  D_\mu\Lambda_2^P - \Lambda_2^N D_\mu \Lambda_1^P) \,, \nonumber\\
  \tilde\Xi_{\mu{\sf a}} &=& g\Lambda^P( T_{P{\sf a}}{}^{\sf b} + 2 d_{{\sf a}
  PN} Z^{N,{\sf b}}) \Xi_{\mu{\sf b}} \,. 
\end{eqnarray}

The field strengths follow from the Ricci identity, $[D_\mu,D_\nu]= -
g \mathcal{F}_{\mu\nu}{}^M\,T_M$, and depend only on the
antisymmetric part of $T_{MN}{}^P$, 
\begin{equation}
  \label{eq:field-strength}
  {\cal  F}_{\mu\nu}{}^M =\partial_\mu A_\nu{}^M -\partial_\nu A_\mu{}^M + g\,
  T_{[NP]}{}^M \,A_\mu{}^N A_\nu{}^P \,.
\end{equation}
Because of the lack of closure expressed by (\ref{Jacobi-X}), they do
not satisfy the Palatini identity,
\begin{equation}
  \label{eq:Palatini}
  \delta\mathcal{F}_{\mu\nu}{}^M = 2\, D_{[\mu}\delta A_{\nu]}{}^M -
  2 g\, T_{(PQ)}{}^M \,A_{[\mu}{}^P \,\delta A_{\nu]}{}^Q\,,
\end{equation}
under arbitrary variations $\delta A_\mu{}^M$. Note that the last term
cancels upon multiplication with the generators $T_M$. The result
(\ref{eq:Palatini}) shows that $\mathcal{F}_{\mu\nu}{}^M$ transforms
under gauge transformations as
\begin{equation}
  \label{eq:delta-cal-F}
  \delta\mathcal{F}_{\mu\nu}{}^M= g\, \Lambda^P T_{NP}{}^M
  \,\mathcal{F}_{\mu\nu}{}^N - 2 g\, Z^{M,{\sf a}} (D_{[\mu}
  \Xi_{\nu]{\sf a}} +d_{{\sf a} PQ} \,A_{[\mu}{}^P\,\delta A_{\nu]}{}^Q)
  \,, 
\end{equation} 
and is therefore not covariant. The standard strategy is therefore to
define modified field strengths,
\begin{equation}
  \label{eq:modified-fs}
  {\cal H}_{\mu\nu}{}^M= {\cal F}_{\mu\nu}{}^M  
  + g\, Z^{M,{\sf a}} \,B_{\mu\nu\,{\sf a}}\;,
\end{equation}
by introducing new tensor fields $B_{\mu\nu\,{\sf a}}$ with suitably
chosen gauge transformation rules, so that covariant results can be
obtained. 

At this point we remind the reader that the invariance transformations
in the rigid case implied that the field strengths $G_{\mu\nu}{}^M$
transform under a subgroup of $\mathrm{Sp}(2n,\mathbb{R})$ (c.f.
(\ref{eq:em-duality})). Our aim is to find a similar symplectric
vector of field strengths so that these transformations are generated
in the non-abelian case as well. This is not possible based on the
variations of the vector fields $A_\mu{}^M$, which will never generate
the type of fermionic terms contained in $G_{\mu\nu\Lambda}$. However,
the presence of the tensor fields enables us to achieve our
objectives, at least in part. Just as in the abelian case, we define
an $\mathrm{Sp}(2n,\mathbb{R})$ vector of field strengths
$\mathcal{G}_{\mu\nu}{}^M$ by
\begin{eqnarray}
  \label{eq:cal-G}
  \mathcal{G}^-_{\mu\nu}{}^\Lambda &=& \mathcal{H}^-_{\mu\nu}{}^\Lambda\,,
  \nonumber\\ 
  \mathcal{G}^-_{\mu\nu\Lambda} &=& F_{\Lambda\Sigma}
  \,\mathcal{H}^-_{\mu\nu}{}^\Sigma - 
  \ft18 F_{\Lambda\Sigma\Gamma} \, \bar \Omega_i{}^{\Sigma}
  \gamma_{\mu\nu} \Omega_j{}^{\Gamma}\,\varepsilon^{ij} \,.  
\end{eqnarray}
Note that the expression for $\mathcal{G}_{\mu\nu\Lambda}$ is the
analogue of (\ref{eq:G-explicit}), with $F_{\mu\nu}{}^\Lambda$
replaced by $\mathcal{H}_{\mu\nu}{}^\Lambda$. 

Following \cite{deWit:2005ub} we introduce the following
transformation rule for $B_{\mu\nu{\sf a}}$ (contracted with
$Z^{M,{\sf a}}$, because only these combinations will appear in the
Lagrangian),
\begin{equation}
  \label{eq:B-transf-0}
   Z^{M,{\sf a}}\, \delta B_{\mu\nu\,{\sf a}} = 2\,Z^{M,{\sf a}}
   (D_{[\mu} \Xi_{\nu]{\sf a}} + d_{{\sf a}\,NP} A_{[\mu}{}^N \delta
   A_{\nu]}{}^P)  - 2\,T_{(NP)}{}^M  \Lambda^P
   \mathcal{G}_{\mu\nu}{}^N    \,,
\end{equation}
where $D_\mu \Xi_{\nu{\sf a}}= \partial_\mu \Xi_{\nu{\sf a}} - g
A_\mu{}^M T_{M{\sf a}}{}^{\sf b} \Xi_{\nu{\sf b}}$ with
$T_{M{\sf a}}{}^{\sf b}= -\Theta_M{}^{\sf c} f_{{\sf c}{\sf a}}{}^{\sf b}$ the
gauge group generator in the adjoint representation of
$\mathrm{G}_{\mathrm{rigid}}$.  With this variation the modified field
strengths (\ref{eq:modified-fs}) are invariant under tensor gauge
transformations. Under the vector gauge transformations we derive the
following result,
\begin{eqnarray}
  \label{eq:delta-G/H}
  \delta \mathcal{G}^-_{\mu\nu}{}^\Lambda&=& - g\,\Lambda^P
  T_{PN}{}^\Lambda \,\mathcal{G}^-_{\mu\nu}{}^N  - g\,\Lambda^P
  T^\Gamma{}_P{}^\Lambda \,
  (\mathcal{G}^-_{\mu\nu} - \mathcal{H}^-_{\mu\nu})_\Gamma \,,
\nonumber\\
  \delta \mathcal{G}^-_{\mu\nu\Lambda} &=& - g\,\Lambda^P
  T_{PN\Lambda} \, \mathcal{G}^-_{\mu\nu}{}^N  -g \,
  F_{\Lambda\Sigma}\,\Lambda^P T^\Gamma{}_P{}^\Sigma\,
  (\mathcal{G}^-_{\mu\nu} - \mathcal{H}^-_{\mu\nu})_\Gamma\,, 
\nonumber\\
  \delta(\mathcal{G}^-_{\mu\nu} - \mathcal{H}^-_{\mu\nu})_\Lambda
  &=&  g \, \Lambda^P(  T^\Gamma{}_{P\Lambda} -T^\Gamma{}_P{}^\Sigma
  \, F_{\Sigma\Lambda})\,
  (\mathcal{G}^-_{\mu\nu} - \mathcal{H}^-_{\mu\nu})_\Gamma\,.
\end{eqnarray}
Hence $\delta\mathcal{G}_{\mu\nu}{}^M=-g\,\Lambda^PT_{PN}{}^M\,
\mathcal{G}_{\mu\nu}^N$, just as the variation of the abelian field
strengths $G_{\mu\nu}{}^M$ in the absence of charges, up to terms
proportional to $\Theta^{\Lambda,{\sf a}}(\mathcal{G}_{\mu\nu}
-\mathcal{H}_{\mu\nu})_\Lambda$. According to \cite{deWit:2005ub}, the latter
terms represent a set of field equations. In that case the last
equation of (\ref{eq:delta-G/H}) expresses the well-known fact that,
under a symmetry, field equations transform into field equations.  As a
result the gauge algebra on these tensors closes according to
(\ref{eq:gauge-commutators}), up to the same field equations.

In order that the Lagrangian (\ref{eq:vlagr-kin}) becomes invariant
under the vector and tensor gauge transformations, we have to make a
number of changes.  First of all, we replace the abelian field
strengths $F_{\mu\nu}{}^\Lambda$ in (\ref{eq:vlagr-kin}) by
$\mathcal{H}_{\mu\nu}{}^\Lambda$, so that
\begin{equation}
  \label{eq:def-cal-G}
  \mathcal{G}_{\mu\nu\,\Lambda} = \mathrm{i} \, 
  \varepsilon_{\mu\nu\rho\sigma}\,
  \frac{\partial\mathcal{L}_{\mathrm{vector}}}
  {\partial{\mathcal{H}}_{\rho\sigma}{}^\Lambda}   \;. 
\end{equation}
Under general variations of the vector and tensor fields we then
obtain the result,
\begin{equation}
  \label{eq:var-L-vector}
  \delta\mathcal{L}_{\mathrm{vector}} = -\mathrm{i}
  \mathcal{G}^{+\mu\nu}{}_\Lambda \Big[ D_\mu\delta A_\nu{}^\Lambda +
  \ft14 g\Theta^{\Lambda{\sf a}} (\delta B_{\mu\nu{\sf a}} - 2d_{{\sf a}
  PQ} A_\mu{}^P \delta A_\nu{}^Q)\Big ] + \mathrm{h.c.}  \,.
\end{equation}
The reader can check that the Lagrangian (\ref{eq:vlagr-kin}) is
indeed invariant under the tensor gauge transformations. Even when we
include the transformations of the scalar and spinor fields, the
Lagrangian is, however, not yet invariant under the vector gauge
transformations. For that it is necessary to introduce the following
universal terms to the Lagrangian \cite{deWit:2005ub}, 
\begin{eqnarray}
  \label{eq:Ltop}   
  {\cal L}_{\rm top} &=&
  \ft18\mathrm{i} g\, \varepsilon^{\mu\nu\rho\sigma}\,
  \Theta^{\Lambda{\sf a}}\,B_{\mu\nu\,{\sf a}} \,
  \Big(2\,\partial_{\rho} A_{\sigma\,\Lambda} + g
  T_{MN\,\Lambda} \,A_\rho{}^M A_\sigma{}^N
  -\ft14g\Theta_{\Lambda}{}^{{\sf b}}B_{\rho\sigma\,{\sf b}}\Big)
  \nonumber\\[.9ex]
  &&{}
  + \ft13\mathrm{i} g\, \varepsilon^{\mu\nu\rho\sigma} T_{MN\,\Lambda}\,
  A_{\mu}{}^{M} A_{\nu}{}^{N}
  \Big(\partial_{\rho} A_{\sigma}{}^{\Lambda}
  +\ft14 gT_{PQ}{}^{\Lambda} A_{\rho}{}^{P}A_{\sigma}{}^{Q}\Big)
  \nonumber\\[.9ex]
  &&{}
  + \ft16 \mathrm{i} g\,\varepsilon^{\mu\nu\rho\sigma}T_{MN}{}^{\Lambda}\,
  A_{\mu}{}^{M} A_{\nu}{}^{N}
  \Big(\partial_{\rho} A_{\sigma}{}_{\Lambda}
  +\ft14 gT_{PQ\Lambda} A_{\rho}{}^{P}A_{\sigma}{}^{Q}\Big)
  \;.
\end{eqnarray}
The first term represents a topological coupling of the antisymmetric
tensor fields with the magnetic gauge fields, and the last two terms
are a generalization of the Chern-Simons-like terms
(\ref{eq:cs-electric}) that we encountered in subsection
\ref{sec:gauge-transformations}. Under variations of the vector and
tensor fields, this Lagrangian varies into (up to total derivative terms)
\begin{equation}
  \label{eq:var-L-top}
  \delta\mathcal{L}_{\mathrm{top}} = \mathrm{i}
  \mathcal{H}^{+\mu\nu\Lambda} \, D_\mu\delta A_{\nu\Lambda}  +
  \ft14 \mathrm{i} g\, \mathcal{H}^{+\mu\nu}{}_\Lambda
  \,\Theta^{\Lambda{\sf a}} (\delta B_{\mu\nu{\sf a}} - 2d_{{\sf a} 
  PQ} A_\mu{}^P \delta A_\nu{}^Q) + \mathrm{h.c.} \,.
\end{equation}
Under the tensor gauge transformations this variation becomes equal to
$(\mathrm{i} g\, \mathcal{H}^{+\mu\nu M}\,\Theta_M{}^{\sf a}\,
D_\mu\Xi_{\nu{\sf a}} + \mathrm{h.c.})$. This expression equals a total
derivative by virtue of (\ref{eq:Z-Theta}) and the Bianchi identity,
\begin{equation}
  \label{eq:Bianchi-D}
  D_{[\mu}{\cal H}_{\nu\rho]}{}^{M} = \ft13g\,Z^{M,{\sf
  a}}\,\mathcal{H}_{\mu\nu\rho{\sf a}} \,,
\end{equation}
where 
\begin{equation}
  \label{eq:H-3}
  \mathcal{H}_{\mu\nu\rho{\sf a}} \equiv 3\, D_{[\mu}
  B_{\nu\rho]\,\alpha} +6 \,d_{\alpha\, NP}\,A_{[\mu}{}^{N} 
  (\partial_{\nu} A_{\rho]}{}^P+ \ft13 g T_{[RS]}{}^{P}
  A_{\nu}{}^{R}A_{\rho]}{}^{S})\,.
\end{equation}
In the above equations, covariant derivatives are defined by
$D_{\mu}{\cal H}_{\nu\rho}{}^{M} =
\partial_{\mu}\mathcal{H}_{\nu\rho}{}^{M} + gA_\mu{}^P T_{PN}{}^M
\mathcal{H}_{\nu\rho}{}^N$ and $D_\rho B_{\mu\nu{\sf a}}=
\partial_\rho B_{\mu\nu\alpha} - g A_\rho{}^M T_{M{\sf a}}{}^{\sf b}
B_{\mu\nu{\sf b}}$. Observe that these derivatives are not fully
covariant in view of (\ref{eq:delta-G/H}) and (\ref{eq:B-transf-0}).
Fully covariantized expressions were presented in \cite{deWit:2007mt}
but are not needed below. The gauge invariance of the total
Lagrangian $\mathcal{L}_{\mathrm{vector}} +
\mathcal{L}_{\mathrm{top}}$, will follow upon including the gauge
transformations of the matter fields \cite{deWit:2005ub}.

As we stressed before, the combined gauge invariance of the vector and
tensor gauge fields ensures that the number of physical degrees of
freedom will not change by the introduction of the magnetic vector
gauge fields and the tensor gauge fields \cite{deWit:2005ub}. The
combined gauge algebra is consistent for the tensor fields upon
projection with the embedding tensor, and as it turns out the action
depends only on those field components. If this were not the case, one
would need to introduce new tensor fields of higher rank
\cite{deWit:2004nw,deWit:2005hv}. Indeed, under variation of the tensor
fields one finds
\begin{equation}
  \label{eq:B-field-eq}
  \delta\mathcal{L}_{\mathrm{vector}} +
  \delta\mathcal{L}_{\mathrm{top}} = -  \ft18
  \mathrm{i}g\,\varepsilon^{\mu\nu\rho\sigma} \, 
  (\mathcal{G}-\mathcal{H})_{\mu\nu\Lambda} \,\Theta^{\Lambda{\sf
  a}}\, \delta B_{\rho\sigma{\sf a}}      \,,
\end{equation}
which shows that the components of the tensor fields that are
projected to zero by multiplication with $\Theta^{\Lambda{\sf a}}$ are
not present in the action. Hence those components can be associated
with an additional gauge invariance. A similar situation arises with
the magnetic gauge fields $A_{\mu\Lambda}$. Under variations of the
gauge fields $A_\mu{}^M$ one derives,
\begin{equation}
  \label{eq:A-field-eq}
  \delta\mathcal{L}_{\mathrm{vector}} +
  \delta\mathcal{L}_{\mathrm{top}} =  \ft12 \mathrm{i}\,
  \varepsilon^{\mu\nu\rho\sigma} \,D_\nu \mathcal{G}_{\rho\sigma}{}^M
  \Omega_{MN} \delta A_\mu{}^N \,,
\end{equation}
up to a total derivative and up to terms that vanish as a result of
the field equation for $B_{\mu\nu\alpha}$.
Substituting~(\ref{eq:Bianchi-D}) we rewrite (\ref{eq:A-field-eq}) as
follows,
\begin{equation}
  \label{eq:A-field-eq-2}
  \delta\mathcal{L}_{\mathrm{vector}} +
  \delta\mathcal{L}_{\mathrm{top}} = \ft12 \mathrm{i}\,
  \varepsilon^{\mu\nu\rho\sigma} \left[- {D}_\nu
  \mathcal{G}_{\rho\sigma\Lambda} \,\delta A_\mu{}^\Lambda + \ft16 g\,
  \mathcal{H}_{\nu\rho\sigma{\sf a}} \,\Theta^{\Lambda{\sf a}} \delta
  A_{\mu\Lambda} \right] \,.
\end{equation}
Because the minimal coupling of the gauge fields is always
proportional to the embedding tensor, the full Lagrangian does not
change under variations of the magnetic gauge fields that are
projected to zero by the embedding tensor component
$\Theta^{\Lambda{\sf a}}$, up to terms that are generated by the
variations of the tensor fields through the `universal' variation,
$\delta B_{\mu\nu{\sf a}}=2\,d_{{\sf a}PQ} A_\mu{}^P \delta
A_\nu{}^Q$.
 
Finally, we have been able to identify yet another independent gauge
invariance which acts only on the tensor fields,
\begin{equation}
  \label{eq:Delta-invariance}
  \Theta^{\Lambda{\sf a}}\delta B_{\mu\nu{\sf a}} \propto
  \Delta^{\Lambda\Sigma\rho}{}_\rho\,
  (\mathcal{G}-\mathcal{H})_{\mu\nu\Sigma} - 6\,
  \Delta^{(\Lambda\Sigma)\rho}{}_{[\rho}\,   
  (\mathcal{G}-\mathcal{H})_{\mu\nu]\Sigma} \,,
\end{equation}
where $\Delta^{\Lambda\Sigma\mu}{}_\nu= \Theta^{\Lambda{\sf a}}
\Delta_{\sf a}{}^{\Sigma\mu}{}_\nu$. 

All these gauge symmetries have a role to play in balancing the
degrees of freedom. In \cite{deWit:2005ub} a precise accounting of all
gauge symmetries was bypassed in the analysis. We note that not all of
them have a bearing on the dynamical modes of the theory as they also
act on fields that play an auxiliary role.

\section{Restoring supersymmetry for non-abelian vector multiplets}
\label{sec:rest-supersymm-non}
\setcounter{equation}{0}
In this section we show how the supersymmetry can be restored in the
presence of a gauging. In this way we will find the generalizations of
the massllike and potential terms of order $g$ and $g^2$,
respectively, which were already exhibited in
(\ref{eq:electric-masslike-potential}) for the case of purely electric
charges. In addition we determine the corresponding changes in the
transformation rules.  The supersymmetry transformations that leave
the action corresponding to (\ref{eq:L-vector}) invariant, were given
in (\ref{eq:susyr}).

Introducing electric and magnetic charges, with a uniform gauge
coupling constant $g$ as before, requires a number of universal changes
of the Lagrangian that were already discussed in the previous
section. In $\mathcal{L}_{\mathrm{matter}}$ we have to covariantize
the derivatives as already discussed in subsection \ref{e+m-charges}.
It is convenient to use the representation
(\ref{eq:lagrangian-pieces}).  With the covariantizations included we
thus have
\begin{equation}
  \label{eq:lagrangian-matter-cov}
  \mathcal{L}_{\mathrm{matter}} = -\mathrm{i} \Omega_{MN}\,
  D_\mu X^M \,D^\mu \bar   X^N + 
  \ft14\mathrm{i}\Omega_{MN}\Big[\bar
  \Omega^{iM} \Slash{D} \Omega_i{}^N
  -\bar \Omega_i{}^M \Slash{D} \Omega^{iN} \Big] \,.  
\end{equation} 
In $\mathcal{L}_{\mathrm{vector}}$ we must replace the abelian field
strengths $F_{\mu\nu}{}^\Lambda$ by the modified field strengths
$\mathcal{H}_{\mu\nu}{}^\Lambda$, defined in
(\ref{eq:modified-fs}). Therefore we replace (\ref{eq:vlagr-kin}) by  
\begin{eqnarray}
  \label{eq:vector-H-cov}
  \mathcal{L}_{\mathrm{vector}}  &=&
  \ft14 \mathrm{i}F_{\Lambda\Sigma}\mathcal{H}^{-\Lambda}_{\mu\nu}
  \mathcal{H}^{-\Sigma\,\mu\nu}
  -\ft1{16} \mathrm{i} F_{\Lambda\Sigma\Gamma}\bar\Omega
  _i^\Lambda\,\gamma^{\mu\nu}  
  \mathcal{H}_{\mu\nu}^{-\Sigma}\,\Omega _j^\Gamma\, \varepsilon^{ij}
      \nonumber\\
  &&
  - \ft1{256}\mathrm{i} N^{\Delta\Omega} \Big(F_{\Delta\Lambda\Sigma}
      \bar\Omega _i{}^\Lambda\gamma_{\mu\nu}\Omega _j{}^\Sigma
  \varepsilon^{ij} \Big) \Big(F_{\Gamma\Xi\Omega} 
  \bar\Omega _k{}^\Gamma\gamma^{\mu\nu}\Omega _l{}^\Xi
  \varepsilon^{kl}\Big)+ \mathrm{h.c.}  \,. 
\end{eqnarray}
Furthermore one includes the Lagrangians (\ref{eq:vlagr-4}),
(\ref{eq:vlagr-Y}) and (\ref{eq:Ltop}), which remain unaltered. Up to
an extension of (\ref{eq:electric-masslike-potential}), whose form we
will establish in this section, we do not expect further
modifications. 

Also the supersymmetry transformation rules acquire a number of
modifications, extending space-time derivatives and field strengths to
covariant ones.  Furthermore one has to take account of the presence
of the new magnetic gauge fields and the tensor fields. However, one
also needs a few additional terms in the transformation rules, whose
form will be established in due course. For the moment we use the
following modified transformation rules, where we also include the
variations of the magnetic gauge fields, which we denote by
$\delta_0$,
\begin{eqnarray}
  \label{eq:susy-1}
  \delta_0 X^{\Lambda} & = & \bar{\epsilon}^i \Omega_i{}^{\Lambda}\,,
  \,\nonumber\\ 
  \delta_0 A_{\mu}{}^{\Lambda} & = & \varepsilon^{ij} \bar{\epsilon}_i
  \gamma_{\mu} \Omega_j{}^{\Lambda} + \varepsilon_{ij}
  \bar{\epsilon}^i \gamma_{\mu} \Omega^{j\Lambda}\,,\nonumber\\ 
  \delta_0 A_{\mu\Lambda} & = & F_{\Lambda\Sigma}\, \varepsilon^{ij}
  \bar{\epsilon}_i \gamma_{\mu} \Omega_j{}^{\Sigma} +
  \bar F_{\Lambda\Sigma}\, \varepsilon_{ij}
  \bar{\epsilon}^i \gamma_{\mu}\Omega^{j\Sigma} \,, \nonumber\\    
  \delta_0\Omega_i{}^\Lambda & = & 2 \Slash{D}X^{\Lambda} \epsilon_i
  + \ft12 \gamma^{\mu \nu} \mathcal{H}^-_{\mu\nu}{}^\Lambda
  \varepsilon_{ij} \epsilon^j + Y_{ij}{}^{\Lambda} 
  \epsilon^j\,, \nonumber\\
  \delta_0 Y_{ij}{}^{\Lambda} & = & 2 \bar{\epsilon}_{(i}
  \Slash{D}\Omega_{j)}{}^{\Lambda} + 2 \varepsilon_{ik} 
  \varepsilon_{jl}\, \bar{\epsilon}^{(k} \Slash{D}\Omega^{l)
  \Lambda}   \,. 
\end{eqnarray}
At this point it is convenient to note that the supersymmetry
variations of the scalar, spinor and vector fields can be written in
the form, 
\begin{eqnarray}
  \label{eq:susy-symplectic}
  \delta_0 X^M & = & \bar{\epsilon}^i \Omega_i{}^M \,,
  \,\nonumber\\ 
  \delta_0 A_{\mu}{}^M& = & \varepsilon^{ij} \bar{\epsilon}_i
  \gamma_{\mu} \Omega_j{}^M + \varepsilon_{ij}
  \bar{\epsilon}^i \gamma_{\mu} \Omega^{j M}\,,\nonumber\\ 
  \delta_0 \Omega_i{}^M & = & 2 \Slash{D}X^M \epsilon_i
  + \ft12 \gamma^{\mu \nu} \mathcal{G}^-_{\mu\nu}{}^M 
  \varepsilon_{ij} \epsilon^j + Z_{ij}{}^M\epsilon^j \,,
\end{eqnarray}
where the fermions $\Omega_i{}^M$, the field strengths
$\mathcal{G}_{\mu\nu}{}^M$, and the quantities $Z_{ij}{}^M$ were
defined in (\ref{eq:symplectic-fermion}), (\ref{eq:cal-G}) and
(\ref{eq:symplectic-Z}), respectively. 

Most of the cancellations required for demonstrating the supersymmetry
of the Lagrangian will still take place when derivatives are replaced
by covariant derivatives. A clear exception arises when dealing with
the commutator of two derivatives, because they will lead to field
strengths upon using the Ricci identity. This situation arises for the
variations of the fermion kinetic term. Furthermore, when establishing
supersymmetry for the more conventional Lagrangians, one makes use of
the Bianchi identity for the field strengths, which no longer applies
to the new field strenghts. Of course, the presence of gauge fields in
the covariant derivatives induces new variations. To investigate these
issues, we first determine the supersymmetry variation of
$\mathcal{L}_{\mathrm{matter}}$ under the transformations given above
(up to total derivatives),
\begin{eqnarray}
  \label{eq:L-matter-variations}
  \delta_0\mathcal{L}_{\mathrm{matter}}&=& 
  \mathrm{i} g\, \Omega_{MQ} T_{PN}{}^Q \,\left[D^\mu\bar X^M \, X^N -
  \bar X^M \, D_\mu X^N +\ft12 \bar\Omega^{iM} 
  \gamma^{\mu}\Omega_i{}^N \right] \,  \delta A_\mu{}^P \nonumber \\
  &&
  -\ft1{2} \mathrm{i} g\, \Omega_{MQ} T_{PN}{}^Q \,\left[\bar X^M\,
  \bar\Omega_i{}^N   \gamma^{\mu\nu} \epsilon^i\,
  \mathcal{H}^-_{\mu\nu}{}^P - \mathrm{h.c.} \right] 
  \nonumber\\
  &&
  +\mathrm{i}\, \Omega_{MN} \,\left[\bar\Omega^{iM}\gamma_\nu
  \epsilon^j\,\varepsilon_{ij} \,D_\mu    \mathcal{G}^{-\mu\nu N}
   - \mathrm{h.c.} \right]  \,,
\end{eqnarray}
where we suppressed variations that involve neither the gauge coupling
constant $g$ nor the (modified) field strengths. These variations will
cancel as before. 

It is now easy to verify that the term of order $g^0$ can be
combined with the result from the variation of
$\mathcal{L}_{\mathrm{vector}} + \mathcal{L}_{\mathrm{top}}$ (c.f.
(\ref{eq:var-L-vector}) and (\ref{eq:var-L-top})),
\begin{equation}
  \label{eq:var-L-vector+top}
  \delta_0(\mathcal{L}_{\mathrm{vector}} +\mathcal{L}_{\mathrm{top}}) 
  = - \mathrm{i} \Omega_{MN} \,
  \mathcal{G}^{-\mu\nu M} \; D_\mu\delta A_\nu{}^N  + \mathrm{h.c.} +
  \cdots   \,.
\end{equation}
Upon using the expressions for $\mathcal{G}_{\mu\nu\Lambda}$ and
$\delta A_{\mu\Lambda}$, the combined result thus leads to a total
derivative plus terms proportional to $D_\mu F_{\Lambda\Sigma}$ and
terms cubic in the fermions. These terms cancel for the abelian theory
with an ordinary derivative and the cancellation proceeds identically
when ordinary derivatives are replaced by covariant ones. Note that
nowhere one needs to use the Bianchi identity. This calculation
confirms the correctness of the transformation rule for the magnetic
gauge fields.  Hence we can now concentrate on the remaining terms of
(\ref{eq:L-matter-variations}), which are the only variations left, up
to terms induced by the variation of the tensor fields which we will
need in due course.

To cancel the order-$g$ terms in (\ref{eq:L-matter-variations}) we
need to add new terms in the transformation rules of
$\Omega_i{}^\Lambda$ and $Y_{ij}{}^\Lambda$. Furthermore new terms to
the Lagrangian are required.  For the case of purely electric charges
these terms are known and the obvious strategy is to simply generalize
these terms. This leads to the expressions, 
\begin{eqnarray}
  \label{eq:order-g}
  \delta_g \Omega_i{}^\Lambda &=& - 2 g\, T_{MN}{}^\Lambda \,\bar X^M
  X^N  \,\varepsilon_{ij} \, \epsilon^j\,,  
  \nonumber \\
  \delta_g Y_{ij}{}^\Lambda &=& -4g\, T_{MN}{}^\Lambda\left[ \bar
  \Omega_{(i}{}^M \epsilon^k \,\varepsilon_{j)k}\,\bar X^N  - 
  \bar\Omega^{kM} \epsilon_{(i}\, \varepsilon_{j)k}\,X^N\right]\,, 
  \nonumber \\
  \mathcal{L}_g &=& -\ft12  \mathrm{i} g \,\Omega_{MQ}
  T_{PN}{}^Q \,\left[ \varepsilon^{ij}\, \bar\Omega_i{}^M 
  \Omega_j{}^P \bar X^N - 
  \varepsilon_{ij} \, \bar\Omega^{iM} \Omega^{jP} X^N \right] \,.    
\end{eqnarray}
In the case of purely electric charges the expression for
$\mathcal{L}_g$ reduces to the first expression of
(\ref{eq:electric-masslike-potential}) upon using
(\ref{eq:derivative-invariance}).

Collecting the new variations proportional to the field strengths that
arise as a result of (\ref{eq:order-g}), we find, using
(\ref{eq:cal-G}), (\ref{eq:fermion-der-invariance}) and
(\ref{eq:lin}),
\begin{equation}
  \label{eq:g-G-variations}
  \delta_g\mathcal{L}_{\mathrm{vector}} + \delta_0\mathcal{L}_g = 
  \ft12 \mathrm{i} g\,\Omega_{MQ} T_{PN}{}^Q\, \bar X^M\,\bar
  \Omega_i{}^N\gamma^{\mu\nu}\epsilon^i \,\mathcal{G}^-_{\mu\nu}{}^P
  + \mathrm{h.c.} \,.
\end{equation}
This term is almost identical to the second term of
(\ref{eq:L-matter-variations}) except that is proportional to
$\mathcal{G}_{\mu\nu}{}^M$ rather than to
$\mathcal{H}_{\mu\nu}{}^M$. However, the combination of these two
terms is cancelled by assigning the following variation to the tensor
fields, 
\begin{equation}
  \label{eq:B-susy-transformation}
  \delta B_{\mu \nu \, {\sf a}} = - 2 t_{{\sf a} M}{}^P\Omega_{PN}
  \left(A_{[\mu}{}^M \,\delta A_{\nu]}{}^N - \bar{X}^M \bar{\Omega}_i{}^N
  \gamma_{\mu\nu} \epsilon^i - X^M \bar\Omega^{iN}
  \gamma_{\mu\nu}\epsilon_i\right)\,. 
\end{equation}

At this point one can verify that all other supersymmetry variations
linear in the gauge coupling constant $g$ vanish. Here one makes use
of the various results derived in subsection \ref{e+m-charges}, and in
particular of (\ref{eq:cov-der-invariance}). What remains are the
order-$g^2$ interactions induced by the order-$g$ transformations of
the spinors, which can be written as,
\begin{equation}
  \label{eq:var-Omega^M-g}
  \delta_g\Omega_i{}^M = -2g\, T_{NP}{}^M \, \bar X^N X^P
  \,\varepsilon_{ij}\, \epsilon^j \,.
\end{equation}
The order-$g^2$ variation follows from $\delta_g \mathcal{L}_g$, and
can be written proportional to the supersymmetry variation $\delta
X^M$ given in (\ref{eq:susy-symplectic}), 
\begin{equation}
  \label{eq:order-g2}
  \delta_g \mathcal{L}_g= - 2\mathrm{i} g^2 \,\Omega_{MQ} T_{NP}{}^Q \,
  \bar X^P \delta X^{[M} \; T_{RS}{}^{N]} \, \bar X^R X^S +
  \mathrm{h.c.}  \,. 
\end{equation}
Using the Lie algebra relation (\ref{eq:closure}), as well as the
relation (\ref{eq:cov-der-invariance}), we can write this in a form
that can be integrated. This reveals that these variations can be 
cancelled by the variation of a scalar potential, corresponding to 
\begin{equation}
  \label{eq:scalar-g2-lagrangian}
  \mathcal{L}_{g^2}= \mathrm{i}g^2\, \Omega_{MN} \, T_{PQ}{}^M X^P
  \bar X^Q  \;  T_{RS}{}^N \bar X^R X^S \,. 
\end{equation}
This expression reduces to (\ref{eq:electric-masslike-potential}) for
purely electric gaugings upon using
(\ref{eq:derivative-invariance}). Observe that the charges
$T_{\Lambda\Sigma\Gamma}$ do not contribute to
(\ref{eq:scalar-g2-lagrangian}), as is well known from previous
constructions. 

Before closing this section we determine the supersymmetry algebra by
evaluating the supersymmetry commutator on $X^M$ and $A_\mu{}^M$
(bearing in mind that the magnetic gauge fields $A_{\mu\Lambda}$ can
be contracted with $\Theta^{\Lambda {\sf a}}$ without loss of
generality). The result for the commuatator takes the following form,
\begin{equation}
  \label{eq:susy-commutator-tm}
  {[\delta(\epsilon_1), \delta(\epsilon_2)]} =
  2(\bar\epsilon_2{}^i\gamma^\mu\epsilon_{1i}+ 
  \bar\epsilon_{2i}\gamma^\mu\epsilon_1{}^i) D_\mu + \delta(\Lambda) +
  \delta(\Xi)  \,,
\end{equation}
where the first term corresponds to a covariant translation (covariant
with respect to vector and tensor gauge transformations), and the
second and third terms denote additional vector and tensor gauge
transformations with parameters,
\begin{eqnarray}
  \label{eq:gauge+Xi}
  \Lambda^M &=& 4\,(\bar X^M \,\bar\epsilon_2{}^i\epsilon_1{}^j\,
  \varepsilon_{ij} 
  + X^M \,\bar\epsilon_{2i}\epsilon_{1j}\,\varepsilon^{ij})\,, 
  \nonumber \\
  \Xi_{\mu {\sf a}} &=& - 2\,d_{{\sf a}NP}(A_\mu{}^N A_\nu{}^P
  +2\,\eta_{\mu\nu} 
   \bar X^N X^P ) (\bar\epsilon_2{}^i\gamma^\nu\epsilon_{1i}+  
  \bar\epsilon_{2i}\gamma^\nu\epsilon_1{}^i)  \,. 
\end{eqnarray}
Here we made use of (\ref{eq:symplectic-invariance}) to close the
commutator on $X^M$. For closing the commutator on $A_\mu{}^M$ we used
the field equations for $Y_{ij}{}^M$ (implying that $Z_{ij}{}^M$ is
pseudo-real), and the field equation for $B_{\mu\nu{\sf a}}$.

This concludes the derivation of supersymmetric vector multiplet
Lagrangians with electric and magnetic gauge charges. In the following
section we will consider the coupling to matter by introducing
hypermultiplets. This will lead to additional contributions to the scalar
potential.

\section{Hypermultiplets}
\label{sec:hypermultiplets}
\setcounter{equation}{0} 
In this section we give a brief description of hypermultiplets and
their gaugings, following the framework of
\cite{DeJaegher:1997ka,deWit:1999fp}. The $n_{\mathrm{H}}$
hypermultiplets are described by $4n_{\mathrm{H}}$ real scalars
$\phi^A$, $2n_{\mathrm{H}}$ positive-chirality spinors
$\zeta^{\bar\alpha}$ and $2n_{\mathrm{H}}$ negative-chirality spinors
$\zeta^\alpha$.  Hence target-space indices $A,B,\ldots$ take values
$1,2,\ldots, 4n_{\mathrm{H}}$, and the indices $\alpha,\beta, \ldots$
and $\bar\alpha,\bar\beta, \ldots$ run from 1 to $2n_{\mathrm{H}}$.
The chiral and antichiral spinors are related by complex conjugation
(so that we have $2n_{\mathrm{H}}$ Majorana spinors) under which
indices are converted according to $\alpha\leftrightarrow \bar\alpha$.

The supersymmetry transformations take the form, 
\begin{eqnarray}
  \label{eq:4dsusy}
  \delta_0\phi^A&=& 2( \gamma^A_{i\bar\alpha} \,\bar\epsilon^i 
  \zeta^{\bar \alpha} +  
  \bar\gamma^{Ai}_{\alpha} \,\bar\epsilon_i \zeta^\alpha
  )\,,\nonumber\\ 
  \delta_0\zeta^\alpha  &=&  V_{A\,i}^\alpha \,\slash{\partial}\phi^A
  \epsilon^i  -\delta\phi^A\, 
  \Gamma_{A}{}^{\!\alpha}{}_{\!\beta}\, \zeta^\beta \,, \nonumber\\  
  \delta_0\zeta^{\bar \alpha}&=& \bar V_A^{i\bar\alpha} 
  \,\slash{\partial}\phi^A\epsilon_i -\delta\phi^A\,
  \bar\Gamma_{A}{}^{\!\bar\alpha}{}_{\!\bar \beta} \,\zeta^{\bar
  \beta}  \,,   
\end{eqnarray}
where $\delta_0$ indicates that the variations refer to zero gauge
coupling constant $g$. Here $\Gamma_A{}^\alpha{}_\beta$ and
$\Gamma_A{}^{\bar\alpha}{}_{\bar\beta}$ are the connections associated
with field-dependent reparametrizations of the fermions of the form
$\zeta^\alpha \to S^\alpha{}_{\!\beta} (\phi)\, \zeta^\beta$, and
$\zeta^{\bar\alpha} \to \bar S^{\bar\alpha}{}_{\!\bar\beta} (\phi)\,
\zeta^{\bar\beta}$. Naturally these reparametrizations act on all
quantities carrying indices $\alpha$ and $\bar\alpha$. The curvatures
$R_{AB}{}^\alpha{}_\beta$ and $R_{AB}{}^{\bar\alpha}{}_{\bar\beta}$
associated with these connections take their values in
$\mathrm{sp}(n_\mathrm{H}) \cong \mathrm{usp}(2n_\mathrm{H};\mathbb{C})$.
The quantities $\gamma^A$ and $V_A$ are
$(4n_\mathrm{H})\times(4n_\mathrm{H})$ complex matrices which play the
role of the quaternionic (inverse) vielbeine of the target space.
They satisfy a pseudo-reality condition specified below.

The Lagrangian takes the following form
\begin{equation}
  \label{eq:4dhlagr1}
  \mathcal{L}_0= -\ft12 g_{AB}\,\partial_\mu\phi^A\partial^\mu\phi^B  
  -G_{\bar \alpha \beta}( \bar\zeta^{\bar \alpha}\Slash{D} \zeta^\beta
  +\bar\zeta^\beta\Slash{D}\zeta^{\bar\alpha}) -\ft14 
  W_{\bar\alpha\beta\bar\gamma\delta}\, \bar \zeta^{\bar \alpha}   
  \gamma_\mu\zeta^{\beta}\,\bar \zeta^{\bar \gamma} 
  \gamma^\mu\zeta^\delta  \,, 
\end{equation}
with covariant derivatives
\begin{equation}
  \label{eq:cov-zeta}
  D_\mu \zeta^\alpha= \partial_\mu \zeta^\alpha + \partial_\mu\phi^A\, 
  \Gamma_{A}{}^{\!\alpha}{}_{\!\beta} \,\zeta^\beta\,, 
  \quad
  D_\mu \zeta^{\bar\alpha}= \partial_\mu \zeta^{\bar \alpha}  +
  \partial_\mu\phi^A\,\bar\Gamma_{A}{}^{\!\bar\alpha}{}_{\bar \beta}
  \,\zeta^{\bar \beta} \,. 
\end{equation}
The tensor $W_{\bar\alpha\beta\bar\gamma\delta}$ is related to the
Riemann curvature $R_{ABCD}$ associated with the target space metric
$g_{AB}$, as well as to the $\mathrm{sp}(n_\mathrm{H})$ curvatures
mentioned above. Observe that the Lagrangian is invariant under the
$\mathrm{U}(1)$ R-symmetry group which acts by chiral transformations
on the fermion fields. The $\mathrm{SU}(2)$ R-symmetry can only be
realized when the target space has an $\mathrm{SU}(2)$ isometry.

The target-space metric $g_{AB}$, the tensors $\gamma^A$, $V_A$ and
the fermionic hermitean metric $G_{\bar\alpha\beta}$ (i.e., satisfying
$(G_{\bar \alpha\beta})^\ast = G_{\bar \beta\alpha}$) are all
covariantly constant with respect to the Christoffel connection and
the connections $\Gamma_{A}{}^{\!\alpha}{}_{\!\beta}$ and
$\Gamma_{A}{}^{\!\bar\alpha}{}_{\!\bar\beta}$. Furthermore we note the
following relations, 
\begin{eqnarray}
  \label{eq:gamma-v} 
  \bar \gamma_{A\alpha}^j \,V_{Bi}^\alpha &=& \gamma_{Bi\bar\alpha}
  \,\bar V^{j\bar \alpha}_A = - \bar \gamma^j_{B\alpha}
  \,V^\alpha_{i A} + \delta^j_i \,g_{AB}\,, \nonumber\\
  \bar V^{i\bar \alpha}_A \, \gamma^A_{j\bar \beta} &=& \delta^i_j\,  
  \delta^{\bar \alpha}_{\,\bar \beta}\,, \nonumber\\
  g^{AB}\, V_{Ai}^\alpha\,V_{Bj}^\beta &=& \varepsilon_{ij} \,
  \Omega^{\alpha\beta}\,,\qquad  g_{AB}\,
  \gamma^A_{i\bar\alpha}\,\gamma^B_{j\bar \beta} = \varepsilon_{ij} \,
  \Omega_{\bar\alpha\bar \beta}\,,  \nonumber\\
  \varepsilon_{ij}\,\Omega_{\bar\alpha\bar\beta}\,\bar
  V^{j\bar\beta}_A &=& g_{AB} 
  \,\gamma^B_{i\bar\alpha} = G_{\bar\alpha\beta}\,V^\beta_{A\,i}\,,
  \nonumber \\
  \gamma_{Ai\bar\alpha}\, \bar V^{j\bar\alpha}_B&=& \varepsilon_{ik}
  J^{kj}_{AB} + \ft12 g_{AB}\, \delta^j_i\,, \nonumber\\ 
  J_{AB}{}^{ij} \gamma^B_{\bar\alpha k} &=& - \delta_k^{(i}
  \,\varepsilon^{j)l} \,\gamma_{A\bar\alpha l}\,. 
\end{eqnarray}
Here $\Omega^{\alpha\beta}$ and $\Omega_{\bar\alpha\bar\beta}$ are
skew-symmetric covariantly constant tensors (satisfying
$\Omega_{\bar\alpha\bar\beta} \bar\Omega^{\bar\beta\bar\gamma}= -
\delta_{\bar\alpha}{}^{\bar\gamma}$), and the $J^{ij}_{AB}$ are
three complex structures generating the algebra of quaternions. The
existence of the complex structures implies that the target space is
hyperk\"ahler.

The equivalence transformations of the fermions and the target-space
diffeomorphisms do not constitute invariances of the theory, unless
they leave the metric $g_{AB}$ and the
$\mathrm{Sp}(n_{\mathrm{H}})\times\mathrm{Sp}(1)$ one-form
$V^\alpha_i$ (and thus the related geometric quantities) invariant.
Therefore invariances are related to isometries of the hyperk\"ahler
space. A subset of them can be elevated to a group of local (i.e.
space-time-dependent) transformations, which require a coupling to
corresponding vector multiplets. Such gauged isometries have been
studied in the literature
\cite{Sierra:1983uh,Hull:1985pq,Bagger:1987rc,D'Auria:1990fj,Andrianopoli:1996cm,deWit:2001bk}
but only for electric charges.

Infinitesimal isometries are characterized by Killing vectors and the
ones associated to local transformations will be labeled by the same
index $M$ that labels the electric and magnetic gauge fields of the
previous sections. In principle, the gauged isometries constitute a
subgroup of the full group of isometries, defined by the embedding
tensor. Hence the corresponding Killing vectors are proportional to
the embedding matrix, $k^A{}_M= \Theta_M{}^{\sf a} \,k^A{}_{\sf a}$,
and (\ref{eq:Z-Theta}) implies, 
\begin{equation}
  \label{eq:Z-killing}
  Z^{M,{\sf a}} \,k^A{}_M =0\,. 
\end{equation}
Without gauge interactions, the hypermultiplets do not couple to the
vector multiplets, so that the full group of invariances factorizes
into separate invariance groups of the vector multiplet Lagrangian and
of the hypermultiplet Lagrangian. The index ${\sf a}$ refers to all
these symmetries, and therefore $k^A{}_{\sf a}$ will vanish whenever
the index ${\sf a}$ refers to a generator acting exclusively on the
vector multiplets.

The local gauge group is thus generated by the Killing vectors
$k^A{}_M(\phi)= (k^A{}_\Lambda(\phi), k^{A\Lambda}(\phi))$, with
parameters $\Lambda^M$. Under infinitesimal transformations we have
\begin{equation}
  \label{eq:delta-phi-Killing} 
\delta\phi^A=g\,\Lambda^M k^A{}_M(\phi)\ ,
\end{equation}
where $g$ is the coupling constant and the $k^A{}_M(\phi)$ satisfy the
Killing equation,
\begin{equation}
  \label{eq:killing-eq}
  D_Ak_{BM} + D_B k_{AM}=0\,. 
\end{equation} 
Higher derivatives of Killing vectors are not independent, as is
shown by
\begin{equation}
  \label{eq:symmetric}
  D_{A}D_{B} k_{C M}  =  R_{BCAE}\,  k^{\,E}{}_M   \ . 
\end{equation}
The isometries close under commutation, 
\begin{equation}
  \label{eq:killingclosure}
  k^B{}_M\partial_Bk^A{}_N-k^B{}_N\partial_Bk^A{}_M = T_{MN}{}^P\,
  k^A{}_P \ ,   
\end{equation} 
where, as before, the antisymmetry in $[MN]$ on the right-hand side is
ensured by (\ref{eq:Z-killing}). 

The invariances associated with the target space isometries act on the
fermions by field dependent matrices, which satify the relation
\begin{equation}
  \label{eq:fermion-t-relation}
  (t_M)^{\alpha}{}_{\!\beta}\, V^\beta_{Ai}  = D_A k^B{}_M\,
  V^\alpha_{Bi}\,, 
\end{equation}
leading to
\begin{equation}
  \label{eq:fermion-t}
  (t_M)^{\alpha}{}_{\!\beta} = \ft12 V_{Ai}^{\alpha} \,
  \bar\gamma^{Bi}_{\beta}\; D_B k^A{}_M\,. 
\end{equation}
The result (\ref{eq:fermion-t-relation}) was derived by requiring that
the tensor $V_{Ai}^\alpha$ is invariant under the isometries, up to a
rotation on the indices $\alpha$.  The invariance implies that
target-space scalars satisfy algebraic identities such as
\begin{equation}
  \label{eq:t-G-Omega}
  \bar t_M{}^{\bar\gamma}{}_{\!\bar\alpha} \, G_{\bar\gamma\beta} 
  +{t_M}^{\gamma}{}_{\!\beta} \, 
  G_{\bar\alpha\gamma}= {t_M}^{\bar\gamma}{}_{\![\bar\alpha} \,
  \Omega_{\bar\beta]\bar\gamma} = 0\,,
\end{equation}
which establishes that the matrices
${t_M}^\alpha{}_\beta$ take values in $\mathrm{sp}(n_{\mathrm{H}})$.
From (\ref{eq:killingclosure}) and (\ref{eq:symmetric}), one may
derive 
\begin{equation}
  \label{eq:t-der}
  D_A t_M{}^{\alpha}{}_{\!\beta}  =  
  R_{AB}{}^{\!\alpha}{}_{\!\beta} \,k^B{}_M \,  , 
\end{equation}
for any infinitesimal isometry. From the group property of the
isometries it follows that the matrices $t_M$ satisfy the commutation
relations,
\begin{equation}
  \label{eq:t-comm}
  [\,t_M ,\,t_N\,]^\alpha{}_{\!\beta}   = -T_{MN}{}^P\, 
  (t_P)^\alpha{}_{\!\beta}  + k^A{}_M\,k^B{}_N\, 
  R_{AB}{}^{\!\alpha}{}_{\!\beta} \,,  
\end{equation}
which takes values in $\mathrm{sp}(n_{\mathrm{H}})$.  This result is
consistent with the Jacobi identity.

The previous results imply that the complex structures $J_{AB}^{ij}$
are invariant under the isometries,
\begin{equation}
  \label{eq:triholo}
  k^C{}_M\, \partial_C J_{AB}^{ij} - 2 \partial_{[A}k^C{}_M
  \,J_{B]C}^{ij} = 0\,,
\end{equation}
implying that the isometries are {\it tri-holomorphic}. From
(\ref{eq:triholo}) one shows that $\partial_A(J^{ij}_{BC}\, k^C{}_M)
-\partial_B(J^{ij}_{AC}\, k^C{}_M)=0$, so that, locally, one can
associate three Killing potentials (or moment maps) $\mu^{ij}{}_M$ to
every Killing vector, according to
\begin{equation}
  \label{eq:tri-Killing-potential}
  \partial_A \mu^{ij}{}_M  = J_{AB}^{ij} \,k^{B}{}_M  \,,
\end{equation}
which determines $\mu^{ij}{}_M$ up to a constant. These constants
correspond to Fayet-Iliopoulos terms. Up to such constants one derives
the equivariance condition,
\begin{equation}
  \label{eq:equivariance}
  J^{ij}_{AB}\,k^A{}_M\,k^B{}_N= T_{MN}{}^{P}\,\mu^{ij}{}_P \ , 
\end{equation}
which implies that the Killing potentials transform covariantly under
the isometries,
\begin{equation}
  \label{k-potential-isometry}
  \delta\mu^{ij}{}_M  = \Lambda^N \,k^A{}_N\,\partial_A
  \mu^{ij}_M  = \Lambda^N \, T_{NM}{}^{P} \,\mu^{ij}{}_P\, . 
\end{equation}

Subsequently we consider the consequences of realizing the isometry
(sub)group generated by the $k^A{}_M$ as a local gauge group. The latter
acts on the hypermultiplet fields in the following way,
\begin{equation}
  \label{eq:fermgaugetr}
  \delta\phi = g\,\Lambda^M\,k^{A}{}_M \,,\qquad
  \delta\zeta^\alpha=g\, \Lambda^M {t_M}^{\alpha}{}_{\!\beta}
  \,\zeta^\beta - \delta\phi^A \Gamma_A{}^{\!\alpha}{}_{\!\beta}
  \,\zeta^\beta \,, 
\end{equation}
where the parameters $\Lambda^M$ are functions of $x^\mu$. The
relevant covariant derivatives are equal to, 
\begin{equation}
  \label{eq:cov-hypermultiplet-der} 
  {\cal D}_\mu \phi^A = \partial_\mu \phi^A - g A_\mu{}^M \,k^A{}_M
  \,, \qquad 
  {\cal D}_\mu\zeta^\alpha =\partial_\mu \zeta^\alpha+
  \partial_\mu\phi^A\,\Gamma_A{}^{\!\alpha}{}_{\!\beta}\, 
  \zeta^\beta -gA_\mu{}^M {t_M}^\alpha{}_{\!\beta}\,\zeta^\beta\, . \; 
\end{equation}
These covariant derivatives must be substituted into the
transformation rules (\ref{eq:4dsusy}) and the Lagrangian
(\ref{eq:4dhlagr1}). The covariance of ${\cal D}_\mu\zeta^\alpha$, 
\begin{equation}
  \label{eq:trasnf-cov-cer}
  \delta{\cal D}_\mu \zeta^\alpha=g\, 
  \Lambda^M t_M{}^{\alpha}{}_{\!\beta}\,{\cal D}_\mu \zeta^\beta -
  \delta\phi^A   
  \Gamma_A{}^{\!\alpha}{}_{\!\beta} \,{\cal D}_\mu\zeta^\beta \,. 
\end{equation}
follows from (\ref{eq:t-der}) and (\ref{eq:t-comm}).

Just as for the vector multiplets, the introduction of the gauge
covariant derivatives to the Lagrangian breaks the supersymmetry of
the Lagrangian. To restore supersymmetry we follow the same procedure
as in section \ref{sec:rest-supersymm-non}. But in this case the
situation is somewhat simpler because the electric and magnetic gauge
fields couple to standard hypermultiplet isometries. This means that
the initial results will coincide with those obtained for electric
gaugings.

Let us first present the variations of the Lagrangian
(\ref{eq:4dhlagr1}) with the proper gauge covariantizations and
determine the supersymmetry variation linear in the gauge coupling
constant $g$ and linear in the fermion fields,
\begin{equation}
  \label{eq:delta-L0}
  \delta\mathcal{L}_0= g \,k_{AM} \Big[\gamma^A_{i\bar\alpha} 
  \,\bar\zeta^{\bar\alpha} \gamma^{\mu\nu}\epsilon^i
  \mathcal{F}^-_{\mu\nu}{}^M 
  + \varepsilon^{ij} \,\bar\Omega_i{}^M
  \Slash{\mathcal{D}}\phi^A \epsilon_j + \mathrm{h.c.}\Big] \,. 
\end{equation}
The first term originates from the fact that the commutator of two
covariant derivatives acquires an extra field strength in the
presence of the gauging, whereas the second term originates from the
variation of the gauge fields in the covariant derivatives of the
scalars. The first term can be cancelled by a supersymmetry variation
of the following new term,
\begin{equation}
  \label{eq:Lagr-g-1}
  \mathcal{L}_g^{(1)} = 2g\, k_{AM} \left[\bar \gamma^{Ai}_{\alpha}
  \varepsilon_{ij}\,{\bar\zeta}^\alpha \Omega^{jM} +
  \gamma^A_{i\bar\alpha} \varepsilon^{ij}\,{\bar\zeta}^{\bar\alpha}
  \Omega_j{}^{M}\right]  \,.
\end{equation}
The variations of this term proportional to the field strength
$\mathcal{G}_{\mu\nu}{}^M$ cancel against the term proportional to
$\mathcal{H}_{\mu\nu}{}^M$ (the field strength
$\mathcal{F}_{\mu\nu}{}^M$ can be replaced by
$\mathcal{H}_{\mu\nu}{}^M$ by virtue of (\ref{eq:Z-killing})) by
adding a new term to the variation (\ref{eq:B-susy-transformation}) of
the tensor fields $B_{\mu\nu{\sf a}}$,
\begin{equation}
  \label{eq:delta-B-hyper}
  \delta B_{\mu\nu{\sf a}}=  - 4\mathrm{i} k^A{}_{{\sf a}} \;  
  \left[ \gamma_{Ai\bar\alpha}  
  \,\bar\zeta^{\bar\alpha} \gamma_{\mu\nu}\epsilon^i -
  \bar\gamma^i_{A\alpha}  
  \,\bar\zeta^{\alpha} \gamma_{\mu\nu}\epsilon_i  \right] \,. 
\end{equation}

Another term in the variation of (\ref{eq:Lagr-g-1}) is proportional
to $X^M$ and its complex conjugate. Their cancellation requires the
following extra variations of the hypermultiplet spinors,
\begin{equation}
  \label{eq:susyferm} 
  \delta\zeta^\alpha = 
  2gX^M \,k^A{}_M V^\alpha_{Ai}\,\varepsilon^{ij}\epsilon_j\,,\qquad  
  \delta\zeta^{\bar \alpha}=  2g{\bar X}^M\,k^A{}_M  {\bar V}^{{\bar
  \alpha}i}_{A}\,\varepsilon_{ij}\epsilon^j\ ,
\end{equation}
and an extra term in the Lagrangian equal to
\begin{equation}
    \label{eq:Lagr-g-2}
  \mathcal{L}_g^{(2)} =
   2g\left[ {\bar X}^M{t_M}^{\!\gamma}{}_{\!\alpha}\,\bar
    \Omega_{\beta\gamma}\,{\bar \zeta}^\alpha\zeta^\beta+ 
    X^M{t_M}^{\!\bar\gamma}{}_{\!\bar\alpha}\,
    \Omega_{\bar\beta\bar\gamma}\,{\bar\zeta}^{\bar\alpha}
    \zeta^{\bar\beta}\right]     \ .
\end{equation}
The remaining variations then take the following form.
\begin{eqnarray}
  \label{eq:delta-0+1+2}
  \delta\mathcal{L}_0 +\delta\mathcal{L}_g^{(1)}
  +\delta\mathcal{L}_g^{(2)} &=& - 2 g\,\partial_A \mu^{ij}{}_M\,
  \bar\Omega_i{}^M \Slash{\mathcal{D}}\phi^A \epsilon_j 
  -2 g\,\partial_A \mu_{ijM} \, 
  \bar\Omega^{iM} \Slash{\mathcal{D}}\phi^A \epsilon^j
  \nonumber\\ 
  &&{}
  - 2g\,\left[\partial_A\mu_{ij\Lambda}\, Y^{ij\Lambda} + 
  \partial_A\mu_{ij}{}^\Lambda\,\bar{F}_{\Lambda\Sigma}\,Y^{ij\Sigma}\right]
  \, \bar\gamma^{Ak}_\alpha\,\bar\epsilon_k\zeta^\alpha 
  \nonumber\\ 
  &&{}
  - 2g\,\left[\partial_A\mu_{ij\Lambda}\, Y^{ij\Lambda} + 
  \partial_A\mu_{ij}{}^\Lambda\,{F}_{\Lambda\Sigma}\,Y^{ij\Sigma}\right]
  \, \gamma^A_{k\bar\alpha}\,\bar\epsilon^k\zeta^{\bar\alpha}\,, 
\end{eqnarray}
where we restricted ourselves to variations linear in the fermion
fields and linear in $g$.

To cancel these variations we must include the following new term to
the Lagrangian,
\begin{eqnarray}
  \label{eq:Lagr-g-3}
  \mathcal{L}_g^{(3)} &=& g\,Y^{ij\Lambda} \left[ \mu_{ij\Lambda}  
    +\ft12 (F_{\Lambda\Sigma}+ \bar
    F_{\Lambda\Sigma})\,\mu_{ij}{}^\Sigma\right] 
    \nonumber\\ 
    &&{}
    -\ft14 g\,\left[ F_{\Lambda\Sigma\Gamma}\,
  \mu^{ij\Lambda} \,\bar\Omega_i{}^\Sigma \Omega_j{}^\Gamma
   + \bar F_{\Lambda\Sigma\Gamma} \,
  \mu_{ij}{}^\Lambda\,\bar\Omega^{i\Sigma} \Omega^{j\Gamma} \right]\,, 
\end{eqnarray}
as well as assign new variations of the fields $\Omega_i{}^\Lambda$
and $Y_{ij}^\Lambda$ of the vector multiplet, 
\begin{eqnarray}
  \label{eq:delta-g-Omega-2}
  \delta_g\Omega_i{}^\Lambda &=& 2\,\mathrm{i}g\, \mu_{ij}{}^\Lambda
  \epsilon^j \,, \nonumber\\
  \delta_g Y_{ij}{}^\Lambda &=& 4\,\mathrm{i} g\,k^{A\Lambda}
  \left[\varepsilon_{k(i}\, 
  \gamma_{j)\bar\alpha A} \bar\epsilon^k\zeta^{\bar\alpha} +
  \varepsilon_{k(i}\, \bar\epsilon_{j)} \zeta^{\alpha} \,
  \bar\gamma^{k}_{\alpha A}   \right]\,. 
\end{eqnarray}
This completes the discussion of all the variations linear in $g$ and
in the fermion fields. The result remains valid for the cubic fermion
variations as well. However, new variations arise in second order in
$g$, by the order-$g$ variations in the order-$g$ terms in the
Lagrangian. Here we have to consider the combined results for the
vector multiplets and the hypermultiplets.  All these variations cancel
against the variation of a scalar potential, corresponding to
\begin{equation}
  \label{eq:hyper-pot}
  \mathcal{L}_{g^2} = -2g^2k^A{}_M \,k^B{}_N  \,g_{AB}\,X^M{\bar
    X^N} - \ft12 g^2 \,N_{\Lambda\Sigma} \;\mu_{ij}{}^\Lambda \,
    \mu^{ij\Sigma}\,.  
\end{equation}

\section{Off-shell structure}
\label{sec:off-shell-structure}
\setcounter{equation}{0}
In the absence of magnetic charges, the vector multiplets constitute
off-shell representations of the $N=2$ supersymmetry algebra and the
tensor fields decouple from the theory. However, on the
hypermultiplets the supersymmetry algebra is only realized up to
fermionic field equations. The situation changes crucially when
magnetic charges are present. In that case there are no longer any off-shell
multiplets and the supersymmetry algebra is only realized when the
fields satisfy the field equations of the hypermultiplet spinors
and of the fields $A_{\mu\Lambda}$, $Y_{ij}{}^\Lambda$ and $B_{\mu\nu
  {\sf a}}$. In this section we discuss how the off-shell closure can
be regained for the vector multiplets when magnetic charges are
switched on. In this discussion the hypermultiplet fields play only an
ancillary role.

We start by introducing $2n$ {\it independent} vector multiplets, associated
with the electric and magnetic gauge fields, $A_\mu{}^\Lambda$ and
$A_{\mu \Lambda}$, and collectively denoted by $A_\mu{}^M$. In the
absence of charges, these fields are subject to the standard off-shell
transformation rules,
\begin{eqnarray}
  \label{eq:susy-M}
  \delta X^M & = & \bar{\epsilon}^i \Omega_i{}^{M}\,,
  \,\nonumber\\ 
  \delta A_{\mu}{}^{M} & = & \varepsilon^{ij} \bar{\epsilon}_i
  \gamma_{\mu} \Omega_j{}^{M} + \varepsilon_{ij}
  \bar{\epsilon}^i \gamma_{\mu} \Omega^{j M}\,,\nonumber\\ 
  \delta\Omega_i{}^M & = & 2 \slash{\partial}X^{M} \epsilon_i
  + \ft12 \gamma^{\mu \nu} F^-_{\mu\nu}{}^M 
  \varepsilon_{ij} \epsilon^j + Y_{ij}{}^{M} 
  \epsilon^j\,, \nonumber\\
  \delta Y_{ij}{}^{M} & = & 2 \bar{\epsilon}_{(i}
  \slash{\partial}\Omega_{j)}{}^{M} + 2 \varepsilon_{ik} 
  \varepsilon_{jl}\, \bar{\epsilon}^{(k}\slash{\partial}\Omega^{l)M} \,. 
\end{eqnarray}
We stress once more that, unlike previously, these $2n$ vector
multiplets are independent. In due course we shall see how to make
contact with the previous description. 

The tensor gauge fields $B_{\mu\nu{\sf a}}$ are assigned to off-shell
tensor multiplets. Just as before, the index ${\sf a}$ labels the
independent continuous symmetries of the theory. These multiplets
consist of scalar fields $L^{ij}{}_{{\sf a}}$, positive chirality
spinors $\varphi^i{}_{\sf a}$ (and their negative chirality conjugates
$\varphi_{i{\sf{a}}}$), tensor gauge fields $B_{\mu\nu{\sf a}}$, and
complex scalars $G_{\sf a}$. However, for reasons explained below, we
complexify the scalars $L^{ij}{}_{{\sf a}}$ by introducing {\it
  complex} scalars $P^{ij}{}_{\sf a}$. These fields transform as
vectors under the $\mathrm{SU}(2)$ R-symmetry, and their pseudo-real
parts are proportional to the fields $L^{ij}{}_{{\sf a}}$,
\begin{equation}
  \label{eq:real-L}
  L^{ij}{}_{{\sf a}}= P^{ij}{}_{{\sf a}} + \varepsilon^{ik}
  \varepsilon^{jl} P_{kl{\sf a}}\,.
\end{equation}
The consistency of this extension is ensured by introducing, at the
same time, the local gauge transformations, $P^{ij}{}_{\sf a}(x)\to
P^{ij}{}_{\sf a}(x) + \mathrm{i}\xi^{ij}{}_{{\sf a}}(x)$, where the
gauge parameters $\xi^{ij}{}_{\sf a}$ are pseudo-real, so that $\xi_{ij{\sf
    a}}= \varepsilon_{ik}\varepsilon_{jl}\, \xi^{kl}{}_{\sf a}$. In
terms of the gauge invariant scalars $L^{ij}{}_{{\sf a}}$ we will
obtain the more conventional formulation of the tensor
multiplet.\footnote{
  We use the notation of \cite{deWit:2006gn}, with the exception of
  the tensor field which is rescaled by a factor 2. Note that the
  precise conventions are crucial for making contact with the tensor
  coupling to the vector multiplets, as employed in this paper (in
  particular, note (\ref{eq:deformation})). }    
The supersymmetry variations of the tensor multiplets are now as
follows,
\begin{eqnarray}
  \label{eq:tensor}
  \delta P_{ij{\sf a}} & = & 2\,\varepsilon_{ik}\varepsilon_{jl}\, 
  \bar\epsilon^{(k}\varphi^{l)}{}_{{\sf a}}\ , \nonumber\\ 
  \delta B_{\mu \nu {\sf a}} & = & \ft12\mathrm{i} \bar\epsilon^i
  \gamma_{\mu\nu}\varphi^j{}_{{\sf a}}\, \varepsilon_{ij} -\ft12 \mathrm{i}
  \bar\epsilon_i \gamma_{\mu\nu} \varphi_{j{\sf a}} \,\varepsilon^{ij}
  \,, \nonumber \\
  \delta \varphi^i{}_{{\sf a}} &=& 
  \slash{\partial} (P^{ij}{}_{{\sf a}} + \varepsilon^{ik}
  \varepsilon^{jl} P_{kl{\sf a}})\,\epsilon_j + 2\,\varepsilon^{ij}
  \Slash{H}_{{\sf a}}\epsilon_j - G_{{\sf a}} \epsilon^i \,, 
  \nonumber  \\   
  \delta G_{{\sf a}} &=&
  -2\,\bar\epsilon_i\slash{\partial}\varphi^i{}_{{\sf a}}\,, 
\end{eqnarray}
where $H^\mu{}_{{\sf a}} = \ft1{2} \mathrm{i}
\varepsilon^{\mu\nu\rho\sigma}\partial_\nu B_{\rho\sigma{\sf
    a}}$. Note that the tensor multiplet fields are thus subject to
two local gauge invariances, 
\begin{equation}
  \label{eq:gauge-xi}
  B_{\mu\nu{\sf a}}(x) \to B_{\mu\nu{\sf a}}(x) +
  2\,\partial_{[\mu}\Xi_{\nu]{\sf a}}(x) \,,  \qquad 
  P_{ij{\sf a}}(x)\to P_{ij{\sf a}}(x) + \mathrm{i}
  \xi_{ij{\sf a}}(x) \,.
\end{equation}
Both these transformations appear in the supersymmetry commutation
relation, which takes the form,
\begin{equation}
  \label{eq:susy-commutator-tv}
  {[\delta(\epsilon_1), \delta(\epsilon_2)]} =
  2(\bar\epsilon_2{}^i\gamma^\mu\epsilon_{1i}+ 
  \bar\epsilon_{2i}\gamma^\mu\epsilon_1{}^i) D_\mu +
  \delta(\Xi)+ \delta(\xi)\,, 
\end{equation}
where the first term denotes the translation (covariantized with
$A_\mu{}^M$ and $B_{\mu\nu{\sf a}}$ dependent terms) and the second
and third one correspond to the transformations (\ref{eq:gauge-xi})
with parameters,
\begin{eqnarray}
  \label{eq:susy-comm-Xi}
  \Xi_{\mu{\sf a}} &=&- 
  \mathrm{i}\,(\bar\epsilon_2{}^i\gamma_\mu\epsilon_{1j} + 
  \bar\epsilon_{2j}\gamma_\mu\epsilon_1{}^i) (P_{ik{\sf
  a}}\varepsilon^{kj} +\varepsilon_{ik}P^{kj}{}_{\sf a}) \,, 
  \nonumber\\
  \xi_{ij{\sf a}} &=&{} 4\mathrm{i}\,(\bar\epsilon_2{}^{k}
  \gamma_\mu\epsilon_{1(i} 
  +\bar\epsilon_{2(i}\gamma_\mu\epsilon_1{}^{k})\,\varepsilon_{j)k}\,
  H^\mu{}_{\sf a} \nonumber\\
  &&{}
  + 2\mathrm{i}\, 
  (\bar\epsilon_2{}^k\gamma^\mu\epsilon_{1(i} 
  +\bar\epsilon_{2(i}\gamma^\mu\epsilon_1{}^k)\,
  \,\partial_\mu P_{j)k{\sf a}} 
  \nonumber \\ 
  &&{}
  -2\mathrm{i}\,(\bar\epsilon_2{}^{(k}\gamma^\mu\epsilon_{1m}
  +\bar\epsilon_{2m}\gamma^\mu\epsilon_1{}^{(k})\,
  \varepsilon_{ik}\varepsilon_{jl}\, \partial_\mu P^{l)m}{}_{\sf a} \,. 
\end{eqnarray}

Now we return to the vector multiplets with a deformation parametrized
by the embedding tensor that couples the vector multiplets to a tensor
multiplet background. The deformation is induced by changing the
field strength tensors {\it and} the auxiliary fields in the
supersymmetry transformation for $\Omega_i$ by
\begin{eqnarray}
  \label{eq:deformation}
  F_{\mu\nu}{}^M &\longrightarrow& \mathcal{H}_{\mu \nu}{}^M =
  \mathcal{F}_{\mu\nu}{}^M + g Z^{M ,{\sf a}} B_{\mu \nu {\sf a}}\,, 
  \nonumber\\
  Y_{ij}{}^M &\longrightarrow& \mathcal{Y}_{ij}{}^M = Y_{ij}{}^M -
  \mathrm{i} g Z^{M ,{\sf a}} P_{ij{\sf a}}\ . 
\end{eqnarray}
Observe that $\mathcal{Y}_{ij}{}^M$ is no longer pseudo-real. Since we
insist on the fact that $\mathcal{H}_{\mu\nu}{}^M$ and
$\mathcal{Y}_{ij}{}^M$ remain gauge invariant with respect to
(\ref{eq:gauge-xi}) we assume the following transformation rules for
$A_\mu{}^M$ and $Y_{ij}{}^M$,
\begin{eqnarray}
  \label{eq:gaay}
  \delta A_\mu{}^M = - g Z^{M,{\sf a}} \Xi_{\mu {\sf a}}\ ,\qquad \delta
  Y_{ij}{}^M = - g Z^{M,{\sf a}} \xi_{ij {\sf a}}\ .
\end{eqnarray}
Subsequently we evaluate the supersymmetry commutator on the vector
multiplet fields acting on $X^M$, $A_\mu{}^M$ and $Y_{ij}{}^M$. For
the moment, we assume non-trivial gaugings, generated by the same
matrices $T_{MN}{}^P$ as before. We thus include order-$g$ corrections
to the supersymmetry variations of $\Omega_i{}^M$ and $Y_{ij}{}^M$.
However, the closure of the supersymmetry commutator is non-trivial in
view of the deformation (\ref{eq:deformation}) and the fact that the
$T_{MN}{}^P$ do not satisfy the Jacobi identity. This will lead to new
contributions to the supersymmetry commutator proportional to the
tensor $Z^{M,{\sf a}}$. The result of an explicit calculation shows
that these contributions can all be absorbed in the transformations
parametrized by $\Xi_{\mu{\sf a}}$ and $\xi_{ij{\sf{a}}}$,
\begin{equation}
  \label{eq:susy-commutator-off-shell-v}
  {[\delta(\epsilon_1), \delta(\epsilon_2)]} =
  2(\bar\epsilon_2{}^i\gamma^\mu\epsilon_{1i}+ 
  \bar\epsilon_{2i}\gamma^\mu\epsilon_1{}^i) D_\mu + \delta(\Lambda) +
  \delta(\Xi)+ \delta(\xi)  \,,
\end{equation}
where the first term corresponds to a covariant translation (covariant
with respect to vector and tensor gauge transformations). The
corresponding parameters are equal to
\begin{eqnarray}
  \label{eq:gauge+Xi+xi}
  \Lambda^M &\!\!=\!\!& 4\,(\bar X^M \,\bar\epsilon_2{}^i\epsilon_1{}^j\,
  \varepsilon_{ij} 
  + X^M \,\bar\epsilon_{2i}\epsilon_{1j}\,\varepsilon^{ij})\,, 
  \nonumber \\[2mm]
  \Xi_{\mu {\sf a}} &\!\!=\!\!& -2\,d_{{\sf a}NP}
  (A_\mu{}^N A_\nu{}^P +2\,\eta_{\mu\nu}\bar X^N X^P) 
  (\bar\epsilon_2{}^i\gamma^\nu\epsilon_{1i}
  + \bar\epsilon_{2i}\gamma^\nu\epsilon_1{}^i)
  \nonumber\\
  &&{}
  - \mathrm{i}\,(\bar\epsilon_2{}^i\gamma_\mu\epsilon_{1j} + 
  \bar\epsilon_{2j}\gamma_\mu\epsilon_1{}^i) (P_{ik{\sf
  a}}\varepsilon^{kj} +\varepsilon_{ik}P^{kj}{}_{\sf a})\,,
  \nonumber\\[2mm]  
  \xi_{ij{\sf a}} &\!\!=\!\!& 4\,d_{{\sf a}NP}
  [X^N\stackrel{\leftrightarrow}{D}_\mu \bar X^P
  -\bar\Omega_l{}^N\gamma_\mu\Omega^{lP} \varepsilon_{k(i}
  (\bar\epsilon_2{}^k \gamma^\mu\epsilon_{1j)} +
  \bar\epsilon_{2j)}\gamma^\mu\epsilon_1{}^k)   \nonumber\\ 
  &&{} -4\,d_{{\sf a}NP} \,\varepsilon^{kl}\,
  \bar\epsilon_{2k}\epsilon_{1l}\,  (2\,
  X^NY_{ij}{}^P-\bar\Omega_i{}^N\Omega_j{}^P)  \nonumber\\ 
  &&{} -4\,d_{{\sf a}NP}\,  \varepsilon_{kl}\,
  \bar\epsilon_{2}{}^k\epsilon_{1}{}^l\,  (2\, \bar
  X^NY^{ijP}-\varepsilon_{im}\varepsilon_{jn}
  \bar\Omega^{mN}\Omega^{nP})  \nonumber\\ 
  &&{} + 4\mathrm{i}\,(\bar\epsilon_2{}^{k}
  \gamma_\mu\epsilon_{1(i} 
  +\bar\epsilon_{2(i}\gamma_\mu\epsilon_1{}^{k})\,\varepsilon_{j)k}\,
  H^\mu{}_{\sf a} \nonumber\\
  &&{}
  + 2\mathrm{i}\, 
  (\bar\epsilon_2{}^k\gamma^\mu\epsilon_{1(i} 
  +\bar\epsilon_{2(i}\gamma^\mu\epsilon_1{}^k)\,
  \,D_\mu P_{j)k{\sf a}} 
  \nonumber \\ 
  &&{}
  -2\mathrm{i}\,(\bar\epsilon_2{}^{(k}\gamma^\mu\epsilon_{1m}
  +\bar\epsilon_{2m}\gamma^\mu\epsilon_1{}^{(k})\,
  \varepsilon_{ik}\varepsilon_{jl}\, D_\mu P^{l)m}{}_{\sf a} 
  \;,
\end{eqnarray}
where use was made of the Bianchi identity (\ref{eq:Bianchi-D}).  The
important observation is that all the terms referring to the tensor
multiplet fields in (\ref{eq:gauge+Xi+xi}) are in precise agreement
(up to the covariantizations) with (\ref{eq:susy-comm-Xi}). The
remaining terms in $\Lambda^M$ and $\Xi_{\mu{\sf a}}$ have already
been found before in (\ref{eq:gauge+Xi}), while those in
$\xi_{ij{\sf{a}}}$ are new.

What remains is to verify the closure on the fermion fields
$\Omega_i{}^M$. In order to do so, we must first extend the tensor
multiplets by incorporating non-abelian gauge couplings in
(\ref{eq:tensor}). However, to keep matters simple, we will suppress
non-abelian gauge interactions here and henceforth. In that case
(\ref{eq:tensor}) is complete and the closure can be verified
directly. As expected, the only possible terms that could affect the
closure are the terms generated by the deformation
(\ref{eq:deformation}). It is then a relatively straightforward
calculation to verify that these terms cancel, so that we have indeed
established the existence of an off-shell representation with both
electric and magnetic charges present. Of course, these charges are
then exclusively carried by hypermultiplets in the way that we have
described before.

Let us now turn to the Lagrangian to see how the on-shell results of
this paper can be obtained. The construction starts from the
observation that, in the absence of the deformations
(\ref{eq:deformation}), there exists a supersymmetric coupling between
tensor and vector supermultiplets. For instance, such a coupling
between the magnetic vector supermultiplets coupling and the tensor
multiplets is described by the following Lagrangian,
\begin{eqnarray}
  \label{eq:initial-L}
  \mathcal{L} &\propto& \Theta^{\Lambda {\sf{a}}}\Big\{
  G_{\sf{a}}X_\Lambda +  \bar G_{\sf{a}}\bar X_\Lambda  -\ft12(
   P_{ij{\sf{a}}} Y^{ij}{}_\Lambda + P^{ij}{}_{\sf{a}} Y_{ij\Lambda})
  \nonumber \\
  &&{}\qquad
  +\bar\Omega^i{}_\Lambda 
  \varphi_{i{\sf{a}}} + \bar\Omega_{i\Lambda} \varphi^i{}_{\sf{a}}
  -\ft12\mathrm{i}\,\varepsilon^{\mu\nu\rho\sigma}
  B_{\mu\nu{\sf{a}}} F_{\rho\sigma\Lambda}    \Big\} \,. 
\end{eqnarray}
In this Lagrangian the tensor multiplet fields act as Lagrange
multipliers which would put the magnetic vector multiplet fields to
zero. Instead, the on-shell theory that we are attempting to construct
should lead to certain relations between the magnetic vector multiplet
fields in terms of the other fields. Moreover, the Lagrangian
(\ref{eq:initial-L}) does not apply in the presence of the
deformations. This suggests to make a number of modifications induced
by the following shifts,
\begin{eqnarray}
  \label{eq:on-shell-subst}
  X_\Lambda &\longrightarrow& X_\Lambda - F_\Lambda(X)\,,\nonumber\\
  \Omega_{i\Lambda} &\longrightarrow& \Omega_{i\Lambda} -
  F_{\Lambda\Sigma}(X)\, \Omega_i{}^\Sigma  \,,\nonumber\\
  F_{\mu\nu\Lambda} &\longrightarrow&
  F_{\mu\nu\Lambda} - \ft14g \Theta_\Lambda{}^{\sf{a}}
  B_{\mu\nu{\sf{a}}} \,,\nonumber\\
  Y_{ij\Lambda} &\longrightarrow& {Y}_{ij\Lambda} +\ft14\mathrm{i}g
  \,\Theta_\Lambda{}^{\sf{a}} P_{ij{\sf{a}}} \,,
\end{eqnarray}
where $F(X)$ is the usual holomorphic function of the scalars
$X^\Lambda$ belonging to the {\it electric} vector multiplets. Note
that the substitutions for $F_{\mu\nu\Lambda}$ and $Y_{ij\Lambda}$
coincide with the expressions for $\mathcal{H}_{\mu\nu\Lambda}$ and
$\mathcal{Y}_{ij\Lambda}$ up to a factor of $\ft12$. This is related
to the fact that the fields $B_{\mu\nu{{\sf{a}}}}$ and
$P_{ij{\sf{a}}}$ will appear quadratically in the Lagrangian upon
performing the shifts (\ref{eq:on-shell-subst}). We note that the
substitution for $Y_{ij\Lambda}$ is ambiguous in view of its
pseudo-reality, while $Y_{ij\Lambda}$ and $Y^{ij}{}_\Lambda$ are
assumed to acquire different shifts.  Ultimately, the justification of
these substitutions is, of course, given by the supersymmetry
invariance of the resulting Lagrangian. Hence without further ado we
now present the following extension of (\ref{eq:initial-L}),
\begin{eqnarray}
  \label{eq:next-L}
  \mathcal{L} &=& - \ft14 g\,\Theta^{\Lambda {\sf{a}}}\Big\{
  G_{\sf{a}}   [X_\Lambda-F_\Lambda(X)]  + \bar G_{\sf{a}}[\bar
  X_\Lambda- \bar F_\Lambda(\bar X)]   \nonumber\\
  &&{}\qquad
  -\ft12\varepsilon^{ik}\varepsilon^{jl}
  P_{ij{\sf{a}}}[Y_{kl}{}_\Lambda +\ft14 \mathrm{i}g
  \Theta_\Lambda{}^{\sf{b}}P_{kl{\sf{b}}}]   
  -\ft12\varepsilon_{ik}\varepsilon_{jl} P^{ij}{}_{\sf{a}}
  [Y^{kl}{}_\Lambda-\ft14\mathrm{i}g\Theta_\Lambda{}^{\sf{b}}
  P^{kl}{}_{\sf{b}}]  \nonumber\\  
  &&{}\qquad
  +\bar\varphi_{i{\sf{a}}}[\Omega^i{}_\Lambda-\bar F_{\Lambda\Sigma}
  \Omega^{i\Sigma}]   
   +\bar\varphi^i_{\sf{a}}[\Omega_{i\Lambda}
  -F_{\Lambda\Sigma}\Omega_i{}^\Sigma]  \nonumber\\
  &&{}\qquad 
  -\ft12\mathrm{i}\,\varepsilon^{\mu\nu\rho\sigma} B_{\mu\nu{\sf{a}}}
  [F_{\rho\sigma\Lambda}- \ft14g \Theta_\Lambda{}^{\sf{b}}
  B_{\rho\sigma{\sf{b}}}] \,\Big\} \,,
\end{eqnarray}
which is invariant under the transformations (\ref{eq:gauge-xi}),
(\ref{eq:gaay}). This Lagrangian is the off-shell extension (in the
abelian case) of (\ref{eq:Ltop}), in view of the fact that the last
term that contains the tensor gauge fields is identical. Clearly, the
fields $G_{\sf{a}}$ and $\varphi^i{}_{\sf{a}}$ act as Lagrange
multipliers, which determine the fields $X_\Lambda$ and
$\Omega_{i\Lambda}$ in the same way as before. The fields
$P^{ij}{}_{\sf{a}}$ can be integrated out, just as the tensor fields
$B_{\mu\nu{\sf{a}}}$.

The Lagrangian (\ref{eq:next-L}) must be combined with the Lagrangian
(\ref{eq:L-vector}) for the electric vector supermultiplets, in which
we have to introduce the deformations (\ref{eq:deformation}). In the
absence of magnetic charges (i.e. $\Theta^{\Lambda{\sf{a}}}=0$), we
thus obtain the standard result for electric charges.

In the presence of magnetic charges, the combined action leads to the
field equation (\ref{eq:B-field-eq}) for the tensor gauge field. For
the field equation associated with $P_{ij}{}^{\sf{a}}$, we should
first exhibit the deformation of the Lagrangian (\ref{eq:vlagr-Y})
that involves the auxiliary fields $Y_{ij}{}^\Lambda$. The correct way
to introduce the deformation reads as follows,
\begin{eqnarray}
  \label{eq:vlagr-Y-deformed}
     \mathcal{L}_Y&=&{}
     -\ft1{8}\mathrm{i}\varepsilon^{ik}\varepsilon^{jl} 
     \Big[F_{\Lambda\Sigma}\,\mathcal{Y}_{ij}{}^\Lambda
     \mathcal{Y}_{kl}{}^\Sigma  -
     F_{\Lambda\Gamma\Omega}\,\mathcal{Y}_{ij}{}^\Lambda 
     \bar\Omega_k{}^\Gamma\Omega_l{}^\Omega \Big]\nonumber\\
     &&{}
     +\ft1{8}\mathrm{i}\varepsilon_{ik}\varepsilon_{jl} 
     \Big[\bar F_{\Lambda\Sigma}\,\mathcal{Y}^{ij\Lambda}
     \mathcal{Y}^{kl\Sigma}  - \bar 
     F_{\Lambda\Gamma\Omega}\,\mathcal{Y}^{ij\Lambda} 
     \bar\Omega^{k\Gamma}\Omega^{l\Omega} \Big]\nonumber\\
     &&{}
     -\ft1{32} N^{\Lambda\Sigma}\Big[\varepsilon^{ik}\varepsilon^{jl}
     F_{\Lambda\Gamma\Omega}\, 
     \bar\Omega_i{}^\Gamma\Omega_j{}^\Omega\; F_{\Sigma\Xi\Delta}
     \,\bar\Omega_k{}^\Xi\Omega_l{}^\Delta +
     \varepsilon_{ik}\varepsilon_{jl} \bar F_{\Lambda\Gamma\Omega} 
     \,\bar\Omega^{k\Gamma}\Omega^{l\Omega} \;
     \bar F_{\Sigma\Xi\Delta} 
     \,\bar\Omega^{i\Xi}\Omega^{j\Delta}\Big] \nonumber\\
     &&{}
     + \ft1{16} N^{\Lambda\Sigma} \,F_{\Lambda\Gamma\Omega}\,
     \bar\Omega_i{}^\Gamma\Omega_j{}^\Omega\; 
     \bar F_{\Sigma\Xi\Delta}\,\bar\Omega^{i\Xi}\Omega^{j\Delta}\,. 
\end{eqnarray} 
Here we observe that the structure of this expression is quite similar
to the structure of (\ref{eq:vlagr-kin}), with the exception of the
last term in the expression above which is separately consistent with
respect to electric/magnetic duality. Actually this term cancels
exactly against the last term of (\ref{eq:vlagr-4}). The result of the
field equations associated with the fields $P_{ij\Lambda}$ can now be
determined and yields,
    \begin{equation}
      \label{eq:P-field-equation}
      \Theta^{\Lambda{\sf{a}}} \Big(\mathcal{Y}_{ij\Lambda}-[
      F_{\Lambda\Sigma} \mathcal{Y}_{ij}{}^\Sigma -\ft12
      F_{\Lambda\Sigma\Gamma} \bar\Omega_i{}^\Sigma\Omega_j{}^\Gamma]
      \Big) = 0\,. 
    \end{equation}
This equation is in close analogy with the field equation
(\ref{eq:B-field-eq}) for the tensor gauge field. 

To make contact with the on-shell results derived in this paper, we
need the terms induced by the gauging for the hypermultiplet
Lagrangian. Starting with the $2n$ independent vector supermultiplets,
the terms of order $g$ and $g^2$ will take the form, 
\begin{eqnarray}
  \label{eq:order-g-Lagrangian-hyper}
  \mathcal{L}_{g+g^2}\Big\vert_{\mathrm{hypermultiplet}} &=&
   +2g\, k_{AM} \left[\bar \gamma^{Ai}_{\alpha}
  \varepsilon_{ij}\,{\bar\zeta}^\alpha \Omega^{jM} +
  \gamma^A_{i\bar\alpha} \varepsilon^{ij}\,{\bar\zeta}^{\bar\alpha}
  \Omega_j{}^{M}\right]  \nonumber\\
   &&{}
   +2g\left[ {\bar X}^M{t_M}^{\!\gamma}{}_{\!\alpha}\,\bar
    \Omega_{\beta\gamma}\,{\bar \zeta}^\alpha\zeta^\beta+ 
    X^M{t_M}^{\!\bar\gamma}{}_{\!\bar\alpha}\,
    \Omega_{\bar\beta\bar\gamma}\,{\bar\zeta}^{\bar\alpha}
    \zeta^{\bar\beta}\right]     \nonumber\\
   &&{}
    +g\,Y^{ijM} \mu_{ijM}  -2g^2k^A{}_M\,k^B{}_N  
    \,g_{AB}\,X^M{\bar X^N}  \,,
\end{eqnarray}
where we include both electric and magnetic Killing potentials. In
principle, one should modify this result by introducing the
deformation (\ref{eq:deformation}). However, the effect of the
deformation drops out in view of the fact that
$Z^{M,{\sf{a}}}\mu_{ijM}=0$, and the hypermultiplet Lagrangian is
separately supersymmetric in the presence of the gauging.

Subsequently we note that the field $Y_{ij\Lambda}$ appears linearly
in the combined Lagrangian, so that it acts as a Lagrange
multiplier. Imposing, at the same time, the gauge condition that
$P_{ij{\sf{a}}}$ is pesudo-real, we obtain the result, 
\begin{equation}
  \label{eq:magn-Y-eq}
  \Theta^{\Lambda{\sf{a}}} \,P_{ij{\sf{a}}} =
  -4 \,\mu_{ij}{}^\Lambda \,. 
\end{equation}
This introduces the correct supersymmetry variation of the fermion
field $\Omega_i{}^\Lambda$, because $\mathcal{Y}_{ij}{}^\Lambda =
Y_{ij}{}^\Lambda + 2 g\mathrm{i}\,\mu_{ij}{}^\Lambda$. Substituting
this last expression into (\ref{eq:vlagr-Y-deformed}) leads then to
additional terms in (\ref{eq:order-g-Lagrangian-hyper}) linear and
quadratic in the magnetic Killing potentials $\mu_{ij}{}^\Lambda$.
These terms coincide with the corresponding terms given in
(\ref{eq:Lagr-g-3}) and (\ref{eq:hyper-pot}).

It should be interesting to further explore the properties and
possible applications of this off-shell formulation. An obvious
question concerns the existence of a non-abelian version. 

\section{Summary and discussion}
\label{sec:summary-discussion}
\setcounter{equation}{0} 
In this paper we presented Lagrangians and supersymmetry
transformations for a general supersymmetric system of vector
multiplets and hypermultiplets in the presence of both electric and
magnetic charges. The results were verified to all orders and are
consistent with results known in the literature that are based on
purely electric charges. The closure of the supersymmetry algebra,
is realized on shell, but in the previous section we have
indicated how an off-shell representation can be defined consisting of
vector and tensor supermultiplets.

Before discussing possible implications of these results, let us first
summarize the terms induced by the gauging. We first present the
combined supersymmetry variations. First of all, we have the original
transformations in the absence of the gauging, where space-time
derivatives are replaced by gauge-covariant derivatives and where the
abelian field strengths $F_{\mu\nu}{}^\Lambda$ are replaced by the
covariant field strengths $\mathcal{H}_{\mu\nu}{}^\Lambda$. We will
not repeat the corresponding expressions here, but we present the
other terms in the transformation rules that are induced by the
gauging. They read as follows,
\begin{eqnarray}
  \label{eq:var-g-fields}
  \delta_g\Omega_i{}^\Lambda &=& -2g\, T_{NP}{}^\Lambda \, \bar X^N X^P
  \,\varepsilon_{ij}\, \epsilon^j + 2\,\mathrm{i}g\, \mu_{ij}{}^\Lambda
  \epsilon^j \,, \nonumber\\ 
  \delta_g\zeta^\alpha &=& 2\,gX^M \,k^A{}_M
  V^\alpha_{Ai}\,\varepsilon^{ij}\epsilon_j\,,\nonumber \\
  \delta_g Y_{ij}{}^\Lambda &=& -4g\, T_{MN}{}^\Lambda\left[ \bar
  \Omega_{(i}{}^M \epsilon^k \,\varepsilon_{j)k}\,\bar X^N  - 
  \bar\Omega^{kM} \epsilon_{(i}\, \varepsilon_{j)k} \,X^N\right]  
  \nonumber \\
  &&{}
  + 4\,\mathrm{i} g\,k^{A\Lambda}
  \left[\varepsilon_{k(i}\, 
  \gamma_{j)\bar\alpha A} \bar\epsilon^k\zeta^{\bar\alpha} +
  \varepsilon_{k(i}\, \bar\epsilon_{j)} \zeta^{\alpha} \,
  \bar\gamma^{k}_{\alpha A}   \right]\,, \nonumber\\
  \delta B_{\mu\nu{\sf a}}&=& - 2 t_{{\sf a} M}{}^P\Omega_{PN}
  \left(A_{[\mu}{}^M \,\delta A_{\nu]}{}^N - \bar{X}^M \bar{\Omega}_i{}^N
  \gamma_{\mu\nu} \epsilon^i - X^M \bar\Omega^{iN}
  \gamma_{\mu\nu}\epsilon_i\right)\nonumber\\
   &&{}
   - 4\mathrm{i} k^A{}_{{\sf a}} \;  
  \left[ \gamma_{Ai\bar\alpha}  
  \,\bar\zeta^{\bar\alpha} \gamma_{\mu\nu}\epsilon^i -
  \bar\gamma^i_{A\alpha}  
  \,\bar\zeta^{\alpha} \gamma_{\mu\nu}\epsilon_i  \right] \,. 
\end{eqnarray}

Likewise we will not repeat the original Lagrangians
(\ref{eq:L-vector}) and (\ref{eq:4dhlagr1}) for the vector multiplets
and hypermultiplets, respectively, modified by the replacement of
space-time derivatives by gauge-covariant ones, and field strengths by
the covariant field strengths $\mathcal{H}_{\mu\nu}{}^\Lambda$. The
Lagrangian (\ref{eq:Ltop}) remains unchanged. The additional terms
induced by the gauging that are linear in $g$ take the following form,
\begin{eqnarray}
  \label{eq:order-g-Lagrangian}
  \mathcal{L}_g &=& -\ft12  \mathrm{i} g \,\Omega_{MQ}
  T_{PN}{}^Q \,\left[ \varepsilon^{ij}\, \bar\Omega_i{}^M 
  \Omega_j{}^P \bar X^N - 
  \varepsilon_{ij} \, \bar\Omega^{iM} \Omega^{jP} X^N \right]
  \nonumber\\
  &&{}
    -\ft14 g\,\left[ F_{\Lambda\Sigma\Gamma}\,
  \mu^{ij\Lambda} \,\bar\Omega_i{}^\Sigma \Omega_j{}^\Gamma
   + \bar F_{\Lambda\Sigma\Gamma} \,
  \mu_{ij}{}^\Lambda\,\bar\Omega^{i\Sigma} \Omega^{j\Gamma} \right]\,, 
    \nonumber\\ 
   &&{}  
   +2g\, k_{AM} \left[\bar \gamma^{Ai}_{\alpha}
  \varepsilon_{ij}\,{\bar\zeta}^\alpha \Omega^{jM} +
  \gamma^A_{i\bar\alpha} \varepsilon^{ij}\,{\bar\zeta}^{\bar\alpha}
  \Omega_j{}^{M}\right]  \nonumber\\
   &&{}
   +2g\left[ {\bar X}^M{t_M}^{\!\gamma}{}_{\!\alpha}\,\bar
    \Omega_{\beta\gamma}\,{\bar \zeta}^\alpha\zeta^\beta+ 
    X^M{t_M}^{\!\bar\gamma}{}_{\!\bar\alpha}\,
    \Omega_{\bar\beta\bar\gamma}\,{\bar\zeta}^{\bar\alpha}
    \zeta^{\bar\beta}\right]     \nonumber\\
   &&{}
    +g\,Y^{ij\Lambda} \left[ \mu_{ij\Lambda}  
    +\ft12 (F_{\Lambda\Sigma}+ \bar
    F_{\Lambda\Sigma})\,\mu_{ij}{}^\Sigma\right] \,.
\end{eqnarray}
The terms of order $g^2$ correspond to a scalar potential proportional
to $g^2$ and are given by
\begin{eqnarray}
  \label{eq:full-g2-lagrangian'}
  \mathcal{L}_{g^2} &=& \mathrm{i}g^2\, \Omega_{MN}\,T_{PQ}{}^M X^P
  \bar X^Q  \;  T_{RS}{}^N \bar X^R X^S \nonumber\\
  &&{}
  -2g^2k^A{}_M \,k^B{}_N  \,g_{AB}\,X^M{\bar
    X^N} - \ft12 g^2 \,N_{\Lambda\Sigma} \,\mu_{ij}{}^\Lambda \,
    \mu^{ij\Sigma}\,.
\end{eqnarray}

Eliminating the auxiliary fields $Y_{ij}{}^\Lambda$ gives rise to the
following expressions. The terms linear in $g$ read, 
\begin{eqnarray}
  \label{eq:Y-L-g}
  \mathcal{L}_g &=& -\ft12  \mathrm{i} g \,\Omega_{MQ}
  T_{PN}{}^Q \,\left[ \varepsilon^{ij}\, \bar\Omega_i{}^M 
  \Omega_j{}^P \bar X^N - 
  \varepsilon_{ij} \, \bar\Omega^{iM} \Omega^{jP} X^N \right]
  \nonumber\\
   &&{}  
   +2g\, k_{AM} \left[\bar \gamma^{Ai}_{\alpha}
  \varepsilon_{ij}\,{\bar\zeta}^\alpha \Omega^{jM} +
  \gamma^A_{i\bar\alpha} \varepsilon^{ij}\,{\bar\zeta}^{\bar\alpha}
  \Omega_j{}^{M}\right]  \nonumber\\
   &&{}
   +2g\left[ {\bar X}^M{t_M}^{\!\gamma}{}_{\!\alpha}\,\bar
    \Omega_{\beta\gamma}\,{\bar \zeta}^\alpha\zeta^\beta+ 
    X^M{t_M}^{\!\bar\gamma}{}_{\!\bar\alpha}\,
    \Omega_{\bar\beta\bar\gamma}\,{\bar\zeta}^{\bar\alpha}
    \zeta^{\bar\beta}\right]     \nonumber\\
   &&{}
    -\ft12\mathrm{i}g N^{\Lambda\Sigma}\,  
    F_{\Sigma\Gamma\Xi}\,
    \,\bar\Omega_i{}^\Gamma\Omega_j{}^\Xi \,\left[\mu^{ij}{}_\Lambda +
    \bar F_{\Lambda\Delta}\,\mu^{ij\Delta} \right] \nonumber\\
    &&{}
    +\ft12\mathrm{i}g N^{\Lambda\Sigma}\,  
    \bar F_{\Sigma\Gamma\Xi}\,
     \Omega^{i\Gamma}\Omega^{j\Xi} \,\left[\mu_{ij\Lambda} +
    F_{\Lambda\Delta}\,\mu_{ij}{}^\Delta \right]   \,.
\end{eqnarray}
The resulting potential, which is proportional to $g^2$, follows from
\begin{eqnarray}
  \label{eq:Y-L-g2}
  \mathcal{L}_{g^2} &=& \mathrm{i}g^2\, \Omega_{MN}\,T_{PQ}{}^M X^P
  \bar X^Q  \;  T_{RS}{}^N \bar X^R X^S 
  -2g^2k^A{}_M \,k^B{}_N  \,g_{AB}\,X^M{\bar
    X^N} \nonumber\\
      &&{}
    -2\,g^2 \left[\mu^{ij}{}_\Lambda +
    F_{\Lambda\Gamma}\,\mu^{ij\Gamma} \right] \,N^{\Lambda\Sigma} 
    \left[\mu_{ij\Sigma} +\bar F_{\Sigma\Xi}\,\mu_{ij}{}^\Xi\right]\,.   
\end{eqnarray}
Provided the embedding tensor is treated as a spurionic quantity, both
these expressions are invariant under electric/magnetic duality
transformations. 

The same phenomenon can be seen in the supersymmetry variation of the
vector multiplet fermions, upon integrating out the fields
$Y_{ij}{}^\Lambda$. Up to terms quadratic in the fermions, this
variation reads, 
\begin{eqnarray}
  \label{eq:delta-Omega}
  \delta\Omega_i{}^\Lambda & = & 2 \Slash{D}X^{\Lambda} \epsilon_i
  + \ft12 \gamma^{\mu \nu} \mathcal{H}^-_{\mu\nu}{}^\Lambda
  \varepsilon_{ij} \epsilon^j  \nonumber\\
  &&{}
  -2g\, T_{NP}{}^\Lambda \, \bar X^N X^P
  \,\varepsilon_{ij}\, \epsilon^j - 4\,gN^{\Lambda\Sigma}(\mu_{ij\Sigma}
  + \bar F_{\Sigma\Gamma} \,\mu_{ij}{}^\Gamma) \epsilon^j \,,
\end{eqnarray}
where the term of order $g$ is consistent with electric/magnetic
duality. 

The above results have many applications. A relatively simple one
concerns the Fayet-Iliopoulos terms, which are the integration
constants of the Killing potentials $\mu^{ij}{}_M$.  This enables us
to truncate the above expressions by setting the embedding tensor to
zero, while still retaining the constants $g\mu^{ij}{}_M$. In that
case all effects of the gauging are suppressed and one is left with a
potential accompanied by fermionic masslike terms,
\begin{eqnarray}
  \label{eq:FI-2}
    \mathcal{L}_{\mathrm{FI}} &=& -\ft12\mathrm{i}g N^{\Lambda\Sigma}\,  
    F_{\Sigma\Gamma\Xi}\,
    \,\bar\Omega_i{}^\Gamma\Omega_j{}^\Xi \,\left[\mu^{ij}{}_\Lambda +
    \bar F_{\Lambda\Delta}\,\mu^{ij\Delta} \right] \nonumber\\
    &&{}
    +\ft12\mathrm{i}g N^{\Lambda\Sigma}\,  
    \bar F_{\Sigma\Gamma\Xi}\,
     \Omega^{i\Gamma}\Omega^{j\Xi} \,\left[\mu_{ij\Lambda} +
    F_{\Lambda\Delta}\,\mu_{ij}{}^\Delta \right] \nonumber\\
    &&{}
    -2\,g^2 \left[\mu^{ij}{}_\Lambda +
    F_{\Lambda\Gamma}\,\mu^{ij\Gamma} \right] \,N^{\Lambda\Sigma} 
    \left[\mu_{ij\Sigma} +\bar F_{\Sigma\Xi}\,\mu_{ij}{}^\Xi\right]
    \,.  
\end{eqnarray}
The above expression transforms as a function under electric/magnetic
duality provided that the $\mu^{ij}{}_M$ are treated as spurionic
quantities transforming as a $2n$-vector under
$\mathrm{Sp}(2n,\mathbb{R})$. To show this one makes use of the
transformation rules (\ref{eq:dual-symm-F-der}) for the second and
third derivatives of the holomorphic function $F(X)$. The last term in
(\ref{eq:FI-2}) corresponds to minus the potential, which is positive
definite (assuming positive $N_{\Lambda\Sigma}$). The Lagrangian is a
generalization of the Lagrangian presented in \cite{Antoniadis:1996},
where it was also shown how the potential can lead to spontaneous
partial supersymmetry breaking when $\mu_{ij}{}^\Lambda\not=0$. Note
that the hypermultiplets play only an ancillary role here, as they
decouple from the vector multiplets.

Most of the possible applications can be found in the context of
supergravity, where they will be useful for constructing low-energy
effective actions associated with string compactifications in the
presence of fluxes (see, e.g. \cite{Grana:2005jc}). In principle it is
straightforward to extend our results to the case of local
supersymmetry. The target space of the vector multiplets should then
be restriced to a special K\"ahler cone (this requires that $F(X)$ be
a homogeneous function of second degree), and the hypermultiplet
scalars should coordinatize a hyperk\"ahler cone. Furthermore the
various formulae for the action and the supersymmetry transformation
rules should be evaluated in the presence of a superconformal
background, so that the action and transformation rules will also
involve the superconformal fields. This has not yet been worked out in
detail for $N=2$ supergravity, although it is in principle
straightforward. In view of the fact that gaugings of $N=4$ and $N=8$
supergravity have already been worked out using the same formalism as
in this paper \cite{Schon:2006kz,deWit:2007mt}, no complications are
expected. Note that Fayet-Iliopoulos terms do not exist in $N=2$
supergravity because the Killing potentials cannot contain arbitrary
integration constants as those would break the scale invariance of the
hyperk\"ahler cone.

The potential is rather independent of all these details, although it
must be rewritten in terms of the appropriate quantities, as was for
instance demonstrated in \cite{deWit:2001bk}. It was already shown in
\cite{deWit:2005ub} that the theory simplifies considerably for
abelian gaugings where $T_{MN}{}^P=0$ and where the potential is
exclusively generated by the hypermultiplet charges. Making use of the
steps described in \cite{deWit:2001bk}, it is rather straightforward
to derive the potential (as was already foreseen in
\cite{deWit:2005ub}), which takes precisely the form conjectured quite
some time ago in \cite{Michelson:1996pn}.

Another application concerns domain wall solutions. In
\cite{Behrndt:2001mx} such solutions were studied in $N=2$
supergravity with both electric and magnetic charges. The
transformation rules postulated in that work are in qualitative
agreement with the ones established in this paper, at least as far as
the terms are concerned that are relevant for the potential (observe
that a magnetic gauge field was absent). A more precise comparison
again requires the extention of our results to the case of local
supersymmetry.

\section*{Acknowledgement}
We thank Stefan Vandoren for valuable discussions. The work of M.d.V.
is part of the research programme of the `Stichting voor Fundamenteel
Onderzoek der Materie (FOM)', which is financially supported by the
`Nederlandse Organisatie voor Wetenschappelijk Onderzoek (NWO)'.\\
This work is also partly supported by NWO grant 047017015, EU contract
MRTN-CT-2004-005104 and INTAS contract 03-51-6346.

\providecommand{\href}[2]{#2}

\end{document}